\documentclass[prd,aps,twocolumn, nofootinbib,]{revtex4}

\usepackage{amsmath,amssymb}
\usepackage{graphicx,psfrag,float,epsfig,color}
\usepackage{mathrsfs,wasysym} \usepackage{epstopdf}
\usepackage{comment}

\DeclareSymbolFontAlphabet{\mathrsfs}{rsfs}
\DeclareMathAlphabet{\mathcal}{OMS}{cmsy}{m}{n}

\newcommand{\be}{\begin{equation}}
\newcommand{\ee}{\end{equation}}

\def\bx{ {\bf x} }
\def\bp{ {\bf p} }

\begin{document}


\title{Caustic echoes from a Schwarzschild black hole}

\author{An{\i}l Zengino\u{g}lu$^1$} \author{Chad R.\!~Galley$^{1,2}$}

\affiliation{$^1$Theoretical Astrophysics, California Institute of
  Technology, Pasadena, California 91125 USA} \affiliation{$^2$Jet
  Propulsion Laboratory, California Institute of Technology, Pasadena,
  California 91109 USA}

\begin{abstract}
We present the first numerical approximation of the scalar Schwarzschild 
Green function in the time domain, which reveals several universal
features of wave propagation in black hole spacetimes. 
We demonstrate the trapping of energy near the photon
sphere and confirm its exponential decay. The trapped wavefront
passes through caustics resulting in echoes that propagate to infinity. 
The arrival times and the decay rate of these
\emph{caustic echoes} are consistent with propagation along null
geodesics and the large $\ell$ limit of quasinormal modes.  We show
that the fourfold singularity structure of the retarded Green
function is due to the well-known action of a Hilbert transform on the
trapped wavefront at caustics. A twofold cycle is obtained for
degenerate source-observer configurations along the caustic line,
where the energy amplification increases with an inverse power of
the scale of the source. Finally, we discuss the tail piece of the solution due
to propagation within the light cone, up to and including null infinity,
and argue that, even with ideal instruments, only a finite number of
echoes can be observed.  Putting these pieces together, we provide a
heuristic expression that approximates the Green function with a few
free parameters. Accurate calculations and approximations of the Green
function are the most general way of solving for wave propagation
in curved spacetimes and should be useful in a variety of studies
such as the computation of the self-force on a particle.
\end{abstract}


\maketitle

\section{Introduction}

A general technique to study the full response
of a black hole to generic perturbations is to explore the Green
function.  While the Green function (aptly called the propagator) is simple 
in flat spacetime, it has a rich structure in curved spacetimes. 

Recently, Ori discovered that the Green function in Schwarzschild
spacetime displays a fourfold periodic structure \cite{Ori}.  This
fourfold structure can be understood in terms of trapped null
geodesics, similar to the classical interpretation of quasinormal mode
ringing as energy leakage from the photon sphere
\cite{AmesThorne:1968, Goebel:1972}. 
At each half revolution along
the trapped orbit, the trapped wavefront passes a caustic undergoing a
transformation with a fourfold cycle.  Ori investigated this cycle
using a quasinormal mode expansion method.  He also
performed an acoustic experiment providing evidence for the structure
of trapped signals through caustics.

The fourfold cycle has been formally analyzed in General Relativity
for the first time by Casals, Dolan, Ottewill, and Wardell in
\cite{Casals:2009xa,Casals:2010zc} in the example of Nariai
spacetimes. Recently, Casals and Nolan developed a Kirchhoff's
integral representation of the Green function on Pleba\'nski--Hacyan
spacetimes, confirming the appearance of the fourfold singularity
structure in these black-hole models \cite{Casals:2012px}.  Going
beyond toy models, Dolan and Ottewill discussed the Green function in
Schwarzschild spacetime using large $\ell$ quasinormal mode sums
\cite{Dolan:2011fh} based on their expansion method \cite{Dolan:2009nk}.
The essential singularity structure of the retarded Green
function in arbitrary spacetimes has been discussed by Harte and
Drivas \cite{Harte:2012uw}. They approximate the region of any spacetime 
near a null geodesic by a pp-wave spacetime in the Penrose limit. Their
analysis implies, in particular, that the fourfold structure should be
valid also in Kerr spacetimes.

Motivated by these theoretical studies and by Ori's acoustic experiment,
we perform a numerical experiment and analyze in detail the
response of a black hole to compactly supported scalar perturbations. 
We describe the problem and our numerical
setup in Sec.~\ref{sec:setup}. The main part of this paper (Sec.~\ref{sec:results}) 
contains our results organized by scales, 
from the geometrical optics limit (short wavelength) 
to the curvature scale (long wavelength). After a
qualitative description of the evolution of compactly supported wave packets in
Sec.~\ref{sec:overview}, we discuss the propagation of the
signal along null geodesics and the subsequent caustic echoes as
measured by an observer at infinity (Sec.~\ref{sec:echoes}).  Wave
propagation through caustics and the fourfold structure of the Green function is studied in Sec.~\ref{sec:hilbert}. We show that the fourfold structure is due to the
action of a Hilbert transform at each passage through a caustic.  In Sec.~\ref{sec:twofold} we present a twofold cycle for degenerate
source-observer configurations and find an inverse power relation between energy
magnification at caustics and the source scale. 
We argue that the trapping of energy near the black hole is the universal cyclic feature manifested in the Green function, not the number of the cycle (e.g., four or two). In Sec.~\ref{sec:tail} we analyze the tail
of the signal due to scattering off the background curvature, which
dominates at late times implying that only a finite number of echoes can be observed even with ideal instruments.  Putting these pieces together in Sec.~\ref{sec:full}, we provide a heuristic expression for the Green
function with a few free parameters. We present our findings, discuss the limitations of our method, and pose problems for future research in Sec.~\ref{sec:conclusions}. The Appendixes complement the main text.  The topics include hyperboloidal compactification for numerical simulations (Appendix \ref{app:layer}), the Schwarzschild Green function in the geometrical optics limit (Appendix \ref{app:optics}), and the Hilbert transform at caustics (Appendix \ref{app:hilbert}).

\section{The Setup}
\label{sec:setup}

\subsection{The problem description}

The simplest model equation to describe essential features of wave
propagation is the scalar wave equation \be\label{eq:sc} \Box\phi(x) =
S(x)\,, \ee for a scalar field $\phi$ with source $S$ that depends on
spacetime coordinates $x$.  The wave operator, $\Box$, is written with
respect to a background Schwarzschild metric. We set the source to
\be\label{eq:source} S(x) = \frac{1}{(2\pi\sigma^2)^2}
\exp\left(-\frac{\delta_{\mu\nu}(x^\mu-x'^\mu)(x^\nu-x'^\nu)}{2\sigma^2}\right),
\ee where $\delta_{\mu\nu}$ is the four-dimensional Kronecker delta
and $\sigma$ sets the scale for the width of the perturbation.  The
Gaussian source to the scalar wave equation is chosen to approximate
the delta distribution because we are interested in the evolution of
wavefronts and in the numerical approximation of the Green
function. The Schwarzschild Green function to the scalar wave equation
satisfies \be\label{eq:green} \Box G(x, x') = - \frac{ 4\pi }{
  \sqrt{-g} } \,\delta^4(x-x') ,\ \ee where $g$ is the metric
determinant, and $\delta^4$ is the four-dimensional Dirac delta
distribution. The solution to the wave equation \eqref{eq:sc} with a
narrow Gaussian source \eqref{eq:source} provides an approximation to
the Schwarzschild Green function with a delta distribution source up
to an overall factor.

\subsection{The numerical setup}

We numerically solve the scalar wave equation \eqref{eq:sc} with
source \eqref{eq:source} using the Spectral Einstein Code {\tt SpEC}
\cite{SpECWebsite}.  {\tt SpEC} is a spectral element code for solving
elliptic and hyperbolic partial differential equations with particular
focus on the Einstein equations for the simulation of compact
binaries.

The time evolution of hyperbolic equations in {\texttt SpEC} uses the
method of lines. A spectral expansion of the unknown in space is
performed in elements that communicate with each other along internal
boundaries through the exchange of characteristics via penalty terms. The
discretized unknown is then evolved as a coupled system of ordinary
differential equations in time with adaptive time-stepping using the
Dormand--Prince method.

\begin{figure}[ht]
\includegraphics[width=\columnwidth]{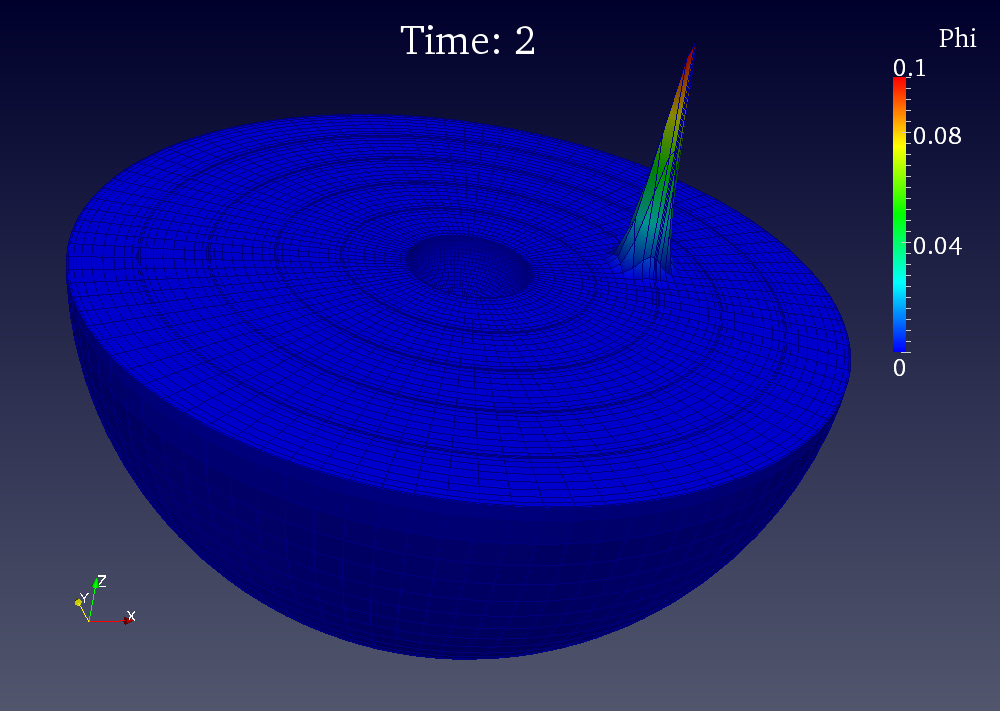}
\caption{Topology of the numerical grid and the initial pulse
  perturbation on the equatorial plane. We only show 5 concentric
  shells of width 2 spanning the radial coordinate domain
  $\rho\in[1.99,12]$ with an expansion into spherical harmonics of
  order 47.  The simulations presented in Sec.~\ref{sec:results}
  use 40 such shells with a spherical harmonic expansion of order 87.
\label{fig:grid}}
\end{figure}

The topology of the grid in {\tt SpEC} can be chosen depending on the
particular problem. We solve the scalar wave equation on a
Schwarzschild background with inner boundary at the event horizon and
outer boundary at null infinity. Therefore, our grid consists of
concentric spherical shells that span the domain between the horizon
and infinity. We use Chebyshev polynomials with Gauss--Lobatto
collocation points in the radial direction and a spherical harmonic
expansion in the angular direction. The topology of the grid and
the initial pulse (\ref{eq:source}) are depicted in Fig.~\ref{fig:grid}.

We employ hyperboloidal scri-fixing \cite{Zenginoglu:2007jw} to
compute the unbounded domain solution and the signal at infinity as
measured by idealized observers.  The coordinates are constructed
using the hyperboloidal layer method \cite{Zenginoglu:2010cq} based on
the ingoing Eddington--Finkelstein coordinates in the interior (Appendix \ref{app:layer}). Setting the Schwarzschild mass $M$, which
provides the scale for the coordinates, to unity, the coordinate domain
in the radial direction is $\rho\in[1.99,12]$ where $\rho=2$
corresponds to the event horizon and $\rho=12$ corresponds to null
infinity.  The interface to the hyperboloidal layer is at $\rho=8$.
The hyperboloidal layer consists of the two outermost shells on the
grid in Fig.~\ref{fig:grid}.

The simulations presented in the results section (Sec.~\ref{sec:results}) employ 40 concentric shells
with 7 collocation points each and a spherical harmonic expansion of
degree 87. For Fig.~\ref{fig:grid} we use a sparse grid for clarity
of visualization.  The Gaussian pulse \eqref{eq:source} is based at
$x'^\mu=\{t',x',y',z'\}=\{2,6,0,0\}$. Our results are robust under variation of parameter values. 

We set the standard deviation as
$\sigma=0.2$ unless otherwise stated, thus giving a 10:1 ratio of the
radius of the event horizon to the scale of the source. 
We would obtain the Green function in the limit of vanishing
width of the Gaussian and infinite numerical resolution. 
Our numerical approximation can be improved using smaller $\sigma$ and higher resolution.

\section{Results}
\label{sec:results}

\subsection{Qualitative description}
\label{sec:overview}

\begin{figure}[t]
\includegraphics[width=\columnwidth]{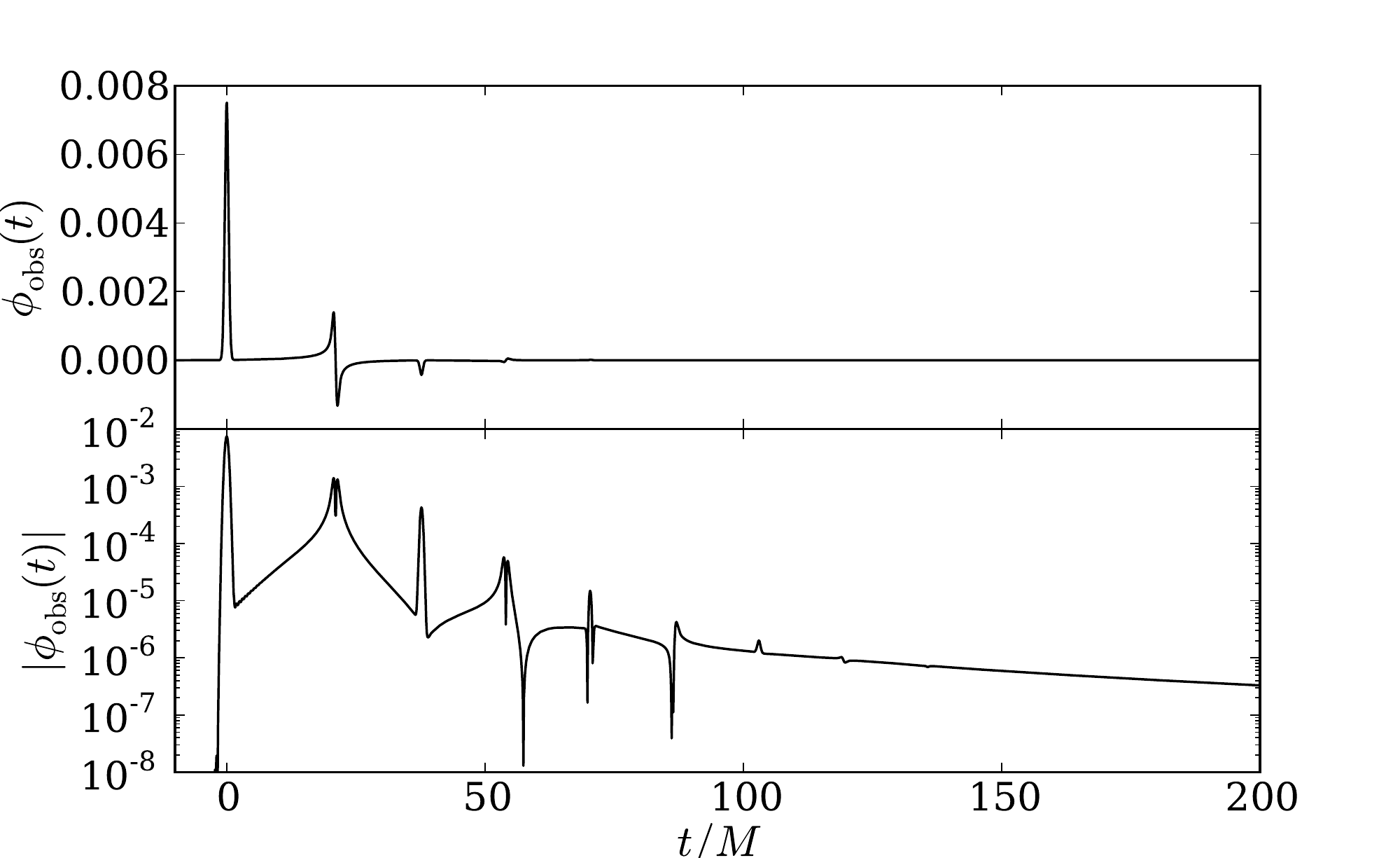}
\caption{{\bf Top:} The evolution of the scalar field generated by a
  compact source at $6M$ on the positive $x$ axis as measured by an
  observer at future null infinity on the positive $z$ axis. {\bf
    Bottom:} The absolute value of the same field on a log scale,
  showing the detailed structure in the amplitude.
\label{fig:logplot}}
\end{figure}

It is well-known that a generic scalar field in Schwarzschild
spacetime dissipates to infinity in an exponentially decaying ringing
and a polynomially decaying tail. The fast dissipation is confirmed in
the top panel of Fig.~\ref{fig:logplot}, which shows the time evolution
of the field as measured by an observer on the positive $z$ axis at null 
infinity.  Following the
initial pulse, we see more structure that becomes visible on a log scale
in the bottom panel. There seems to be a recurring signal caused by
the initial perturbation that disappears at late times when the
polynomial decay dominates.

\begin{figure}[ht]
\includegraphics[width=0.93\columnwidth]{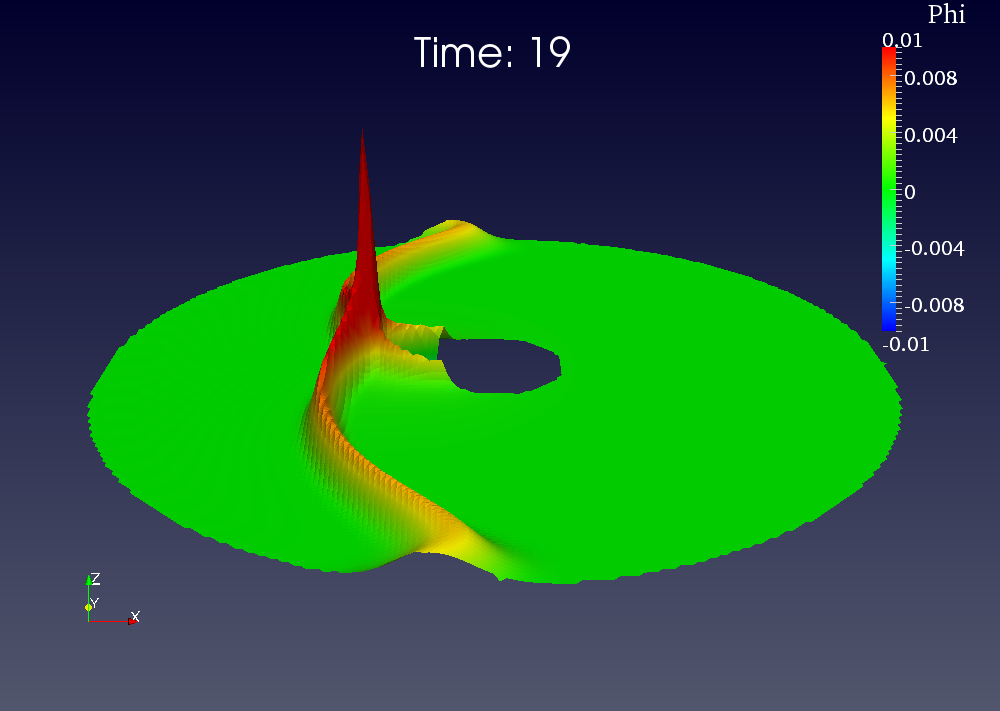}
\caption{Formation of the first caustic echo from the initial pulse in
  Fig.~\ref{fig:grid}. We plot a generic bird's eye view onto the
  equatorial plane. A movie of the dynamical evolution, including a
  top to bottom view, can be seen online \cite{video}. 
  The color and height of the surface is determined
  by the amplitude of the scalar field $\phi$. The wavefront of the
  direct signal has a Gaussian profile and forms a cardioid shape
  around the event horizon, which is typical of caustic formation. The amplitude
  amplification is clearly visible. The first caustic echo forms in
  the wake of the caustic and is seen as a wavefront in the vicinity
  of the horizon. It generates a second caustic echo plotted in
  Fig.~\ref{fig:3D_2}.
\label{fig:3D_1}}
\end{figure}

\begin{figure}[ht]
\includegraphics[width=0.93\columnwidth]{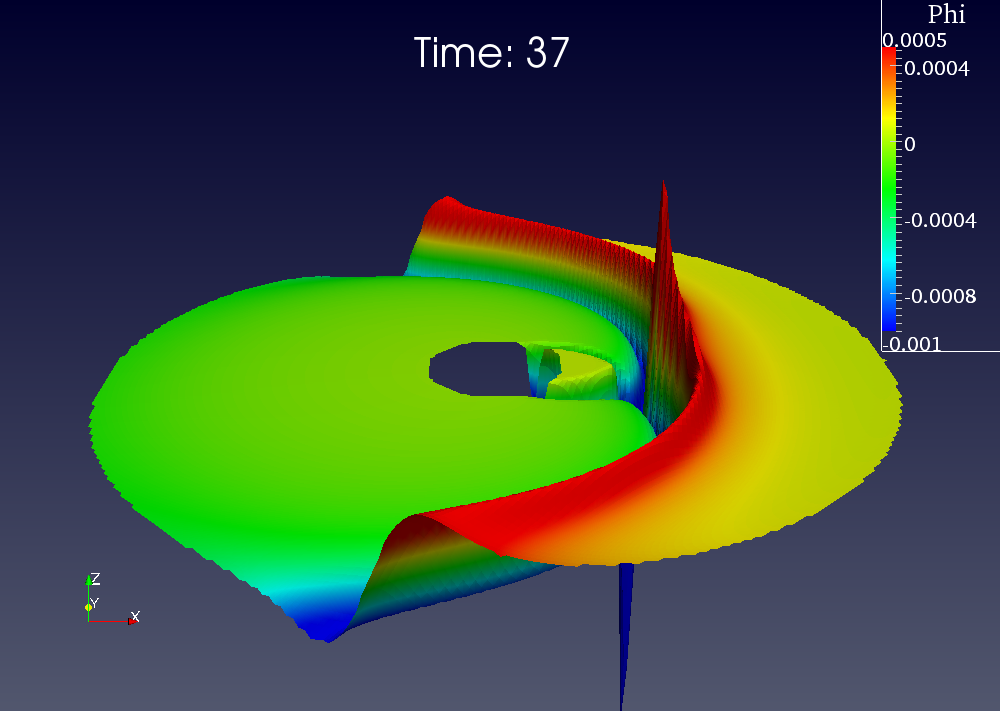}
\caption{Formation of the second caustic echo. The profile of the
  signal from the wavefront of the first echo, seen at the
  boundary, is not a Gaussian but its Hilbert transformation
  (see Sec.~\ref{sec:hilbert}). The wavefront of the second caustic echo
  near the horizon is strictly negative and has a negative Gaussian
  profile (see boundary profile in Fig.~\ref{fig:3D_3}).
\label{fig:3D_2}}
\end{figure}

\begin{figure}[t]
\includegraphics[width=0.93\columnwidth]{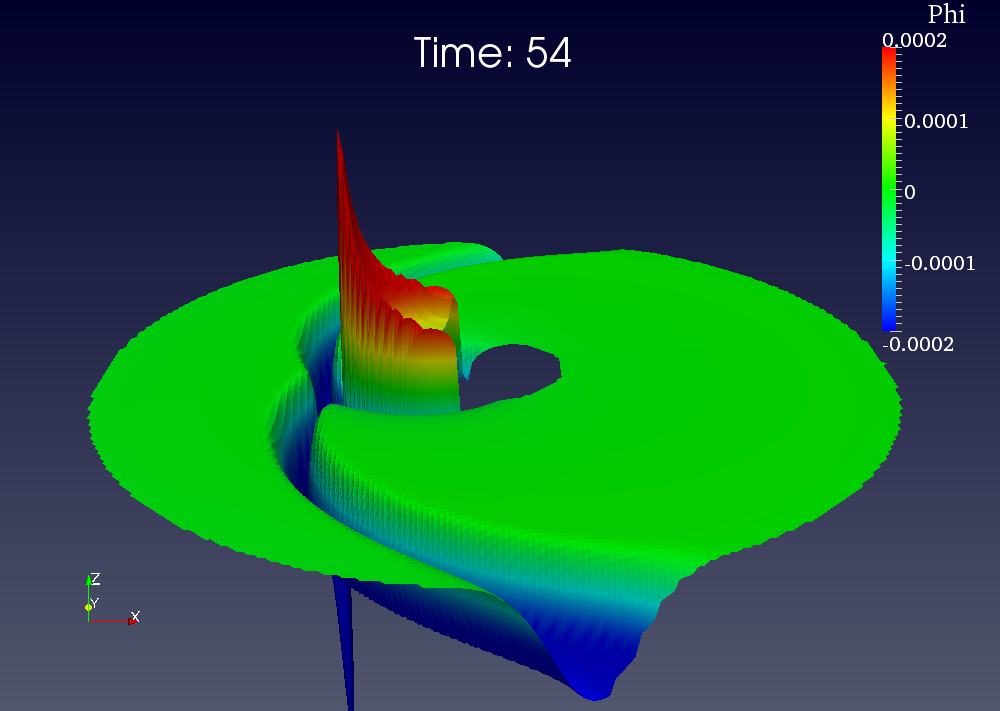}
\caption{Formation of the third caustic echo. The second caustic echo
  has a negative Gaussian profile as seen at the boundary. The profile
  of the third echo is clearer in Fig.~\ref{fig:3D_4}.
\label{fig:3D_3}}
\end{figure}

\begin{figure}[t]
\includegraphics[width=0.93\columnwidth]{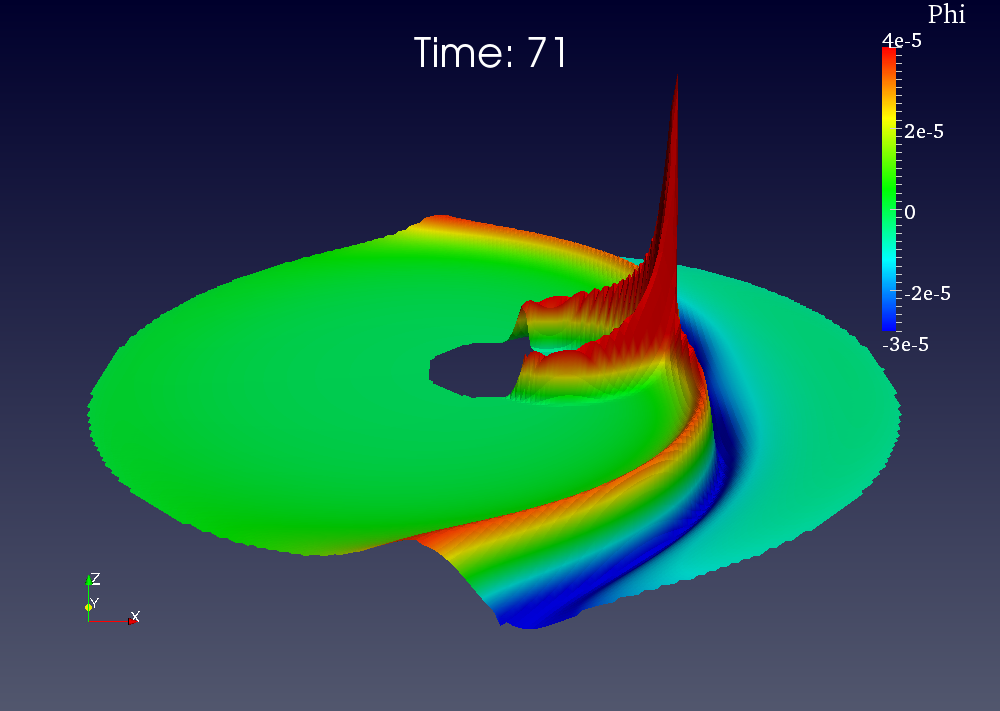}
\caption{Formation of the fourth caustic echo. The profile of the
  third caustic echo is the Hilbert transform of the negative
  Gaussian, which is the negative of the profile of the first
  caustic echo seen in Fig.~\ref{fig:3D_2}. The fourth caustic echo
  emanating from the horizon has a positive Gaussian profile like the
  direct signal, thus completing the fourfold cyle.
\label{fig:3D_4}}
\end{figure}

The dynamics behind Fig.~\ref{fig:logplot} can be seen more clearly in
three-dimensional plots. In Figs.~\ref{fig:3D_1}--\ref{fig:3D_4} 
we show snapshots from the evolution of $\phi$ on the
equatorial plane.  We set the width of the Gaussian source to
$\sigma=0.3M$ for visualization.

The curvature of spacetime near the black hole causes the wavefront
triggered by the initial pulse perturbation in Fig.~\ref{fig:grid} to
bend and to focus forming a half-cardioid shape typical of caustic
formation (Fig.~\ref{fig:3D_1}). At the cusp of the cardioid, by the
antipodal point to the initial perturbation near the black hole, the
wavefront passes through itself thus forming a caustic.  The caustic moves out to
infinity along the negative $x$ axis and leaves an echo in its
wake. This secondary wavefront again forms a cardioid shape with the
cusp on the other side of the black hole (Fig.~\ref{fig:3D_2}). This
process repeats itself resulting in echoes that reach
observers in regular time intervals (Figs.~\ref{fig:3D_3} and
\ref{fig:3D_4}).

The figures suggest that the echoes are due to partial trapping of the
initial energy at the photon sphere at areal coordinate $3M$. 
Part of the wavefront from the
initial perturbation dissipates directly to infinity, part of it falls
into the black hole, and part is trapped near the horizon forming
so-called ``surface waves" propagating on the photon sphere
\cite{AmesThorne:1968, Goebel:1972, Andersson:1994rm,
  Decanini:2010fz}.  The trapped signal goes through two caustics upon
each full revolution and leaks out energy to infinity in the form of
propagating wavefronts that we call {\it caustic echoes}.

The profiles of the caustic echoes follow a certain pattern. The
direct signal has the profile of a Gaussian in accordance with the
source (Figs.~\ref{fig:grid} and \ref{fig:3D_1}). The shape of the first caustic
echo (see the boundary profile in Fig.~\ref{fig:3D_2}) is the Hilbert transform 
of the direct Gaussian signal, which we refer to as a Dawsonian, and is
discussed in more detail in Sec.~\ref{sec:hilbert}. The second echo is
a negative Gaussian (Fig.~\ref{fig:3D_3}) and finally the fourth echo
is a negative Dawsonian (Fig.~\ref{fig:3D_4}). The wavefront
emanating from the wake of the fourth echo has a Gaussian profile just
like the initial signal, thereby completing the fourfold cycle
(Fig.~\ref{fig:3D_4}).

This is the demonstration of the fourfold structure first discovered
in General Relativity by Ori \cite{Ori} and analyzed by Casals
\emph{et} al.~\cite{Casals:2009xa,Casals:2010zc}. A visualization of the
time evolution can be found on the internet \cite{video}. In the following,
we present a detailed quantitative understanding of the data presented
in Fig.~\ref{fig:logplot}. We present analysis for the
measurements of an observer at future null infinity on the $z$ axis,
unless stated otherwise.

\subsection{Arrival and decay of caustic echoes}
\label{sec:echoes}

\begin{figure}[ht]
	\includegraphics[width=\columnwidth]{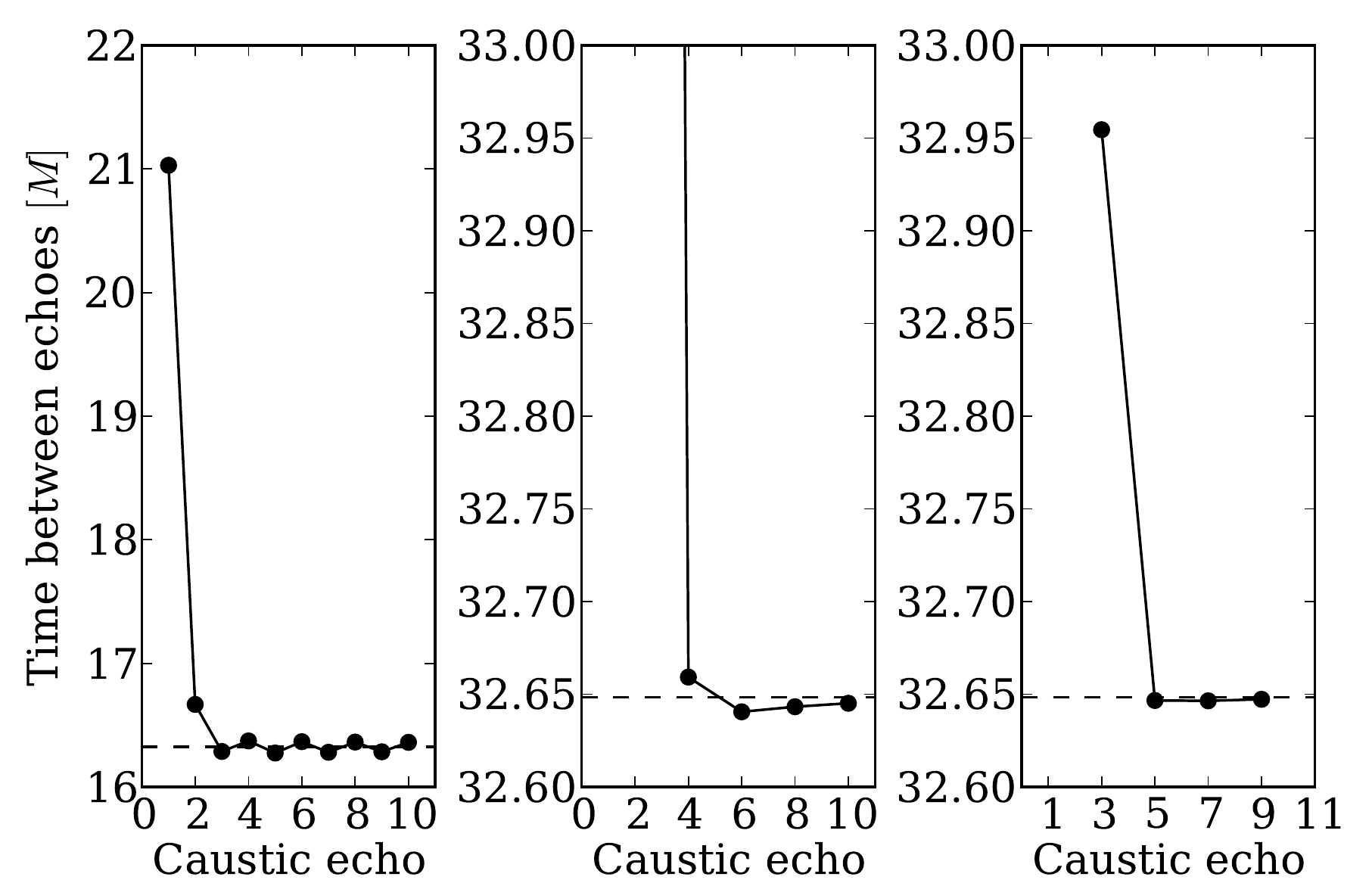}
	\caption{Time differences between caustic echoes indicating agreement with theoretical prediction based on null geodesic revolution times around the photon sphere at $3M$. Our measurement of the arrival times for differently shaped echoes leads to an artificial, systematic deviation from the theoretical prediction ($\pi \sqrt{27}M$, dashed line), which averages out (left panel). When similarly shaped echoes are compared, however, the measured time differences  (between even echoes in the middle panel and between odd echoes in the right panel) agree accurately with the revolution time of null geodesics around the photon sphere given by $T_{{\rm full}}=2\pi \sqrt{27}M \approx 32.65 M$.}
	\label{fig:period1}
\end{figure}

The first feature seen by the observer in Fig.~\ref{fig:logplot} is a narrow Gaussian pulse
that we refer to as the {\it direct signal}, which is the part of the perturbation that propagates directly to the observer.  We set the observer's
clock to zero by the arrival time of the direct signal's maximum. The
arrival times of the subsequent caustic echoes are plotted in
Fig.~\ref{fig:period1}.

Even and odd echoes, that is, echoes resulting from the wavefront
passing through an even or odd number of caustics, have different
profiles (see Figs.~\ref{fig:logplot}-\ref{fig:3D_4} and Sec.~\ref{sec:hilbert}).  Therefore, it is appropriate to consider the
arrival times of even and odd echoes separately.  We extract the times
between even echoes by fitting a Gaussian and reading off the time at
which the peak amplitude arrives at the observer (middle panel of
Fig.~\ref{fig:period1}).  We extract the times between odd echoes by
fitting their time derivative to a Gaussian and finding the time at
which the peak of the derivative is maximal (right panel of
Fig.~\ref{fig:period1}).

The arrival times of caustic echoes approach a constant value. This
observation is explained through the generation mechanism behind the
echoes, namely, the revolution of trapped energy at the photon sphere along
null geodesics. The quantitative prediction of the arrival times
requires only the study of high-frequency wave propagation for which
it suffices to employ the geometrical optics approximation.  An
analysis of solutions to the scalar wave equation in Schwarzschild
spacetime with localized energy has been performed by Stewart
\cite{Stewart:1989}. Stewart shows that the peak amplitude propagates
along null geodesics, which orbit the photon sphere 
with a period of $T_{{\rm full}}=2\pi \sqrt{27}M \approx 32.6484 M$. 
The half-period for the formation of each successive caustic is then 
$T_{{\rm half}}=\pi \sqrt{27} M\approx 16.3242 M$. 
These values are plotted as dashed lines
in Fig.~\ref{fig:period1} and indicate good agreement of the measured
arrival times with the theoretical prediction. The best-fit value for
the successive echoes shown in the left panel of
Fig.~\ref{fig:period1} agrees with $T_{{\rm half}}$ to 4 significant
digits.

The geometrical optics approximation is a short-wave (or
high-frequency) approximation, and should therefore agree with the
large $\ell$ limit of quasinormal modes. In this limit, the
quasinormal mode frequencies can be expressed analytically as
\cite{Ferrari:1984zz, Iyer:1986np}
\begin{align}
	M \omega^{m=\ell}_{\ell n} \sim \frac{1}{\sqrt{27}} \big( \ell
        - i (n+1/2) \big)  .
\label{eq:largeLqnm1}
\end{align}
The real part of the frequency (divided by $\ell$) corresponds to the
full orbital frequency $2\pi / T_{\rm full}$ of null rays on the unstable photon
orbit. Thus, the time between successive even or odd caustic echoes is
related to the real part of quasinormal mode frequencies in the large
$\ell$ limit, which in turn are related to properties of null
geodesics at or near the unstable photon orbit.  

From
\eqref{eq:largeLqnm1} it is clear that the amplitude of caustic
echoes should decay exponentially with a decay rate given by the
imaginary part of the large $\ell$ frequency for the lowest overtone
(hence longest decay time), $n=0$. The same decay rate can be
calculated from the geometrical optics limit \cite{Stewart:1989} and
is given by the Lyapunov exponent of unstable photon orbits
\cite{Goebel:1972, Cardoso:2008bp}.

\begin{figure}[ht]
	\includegraphics[width = \columnwidth]{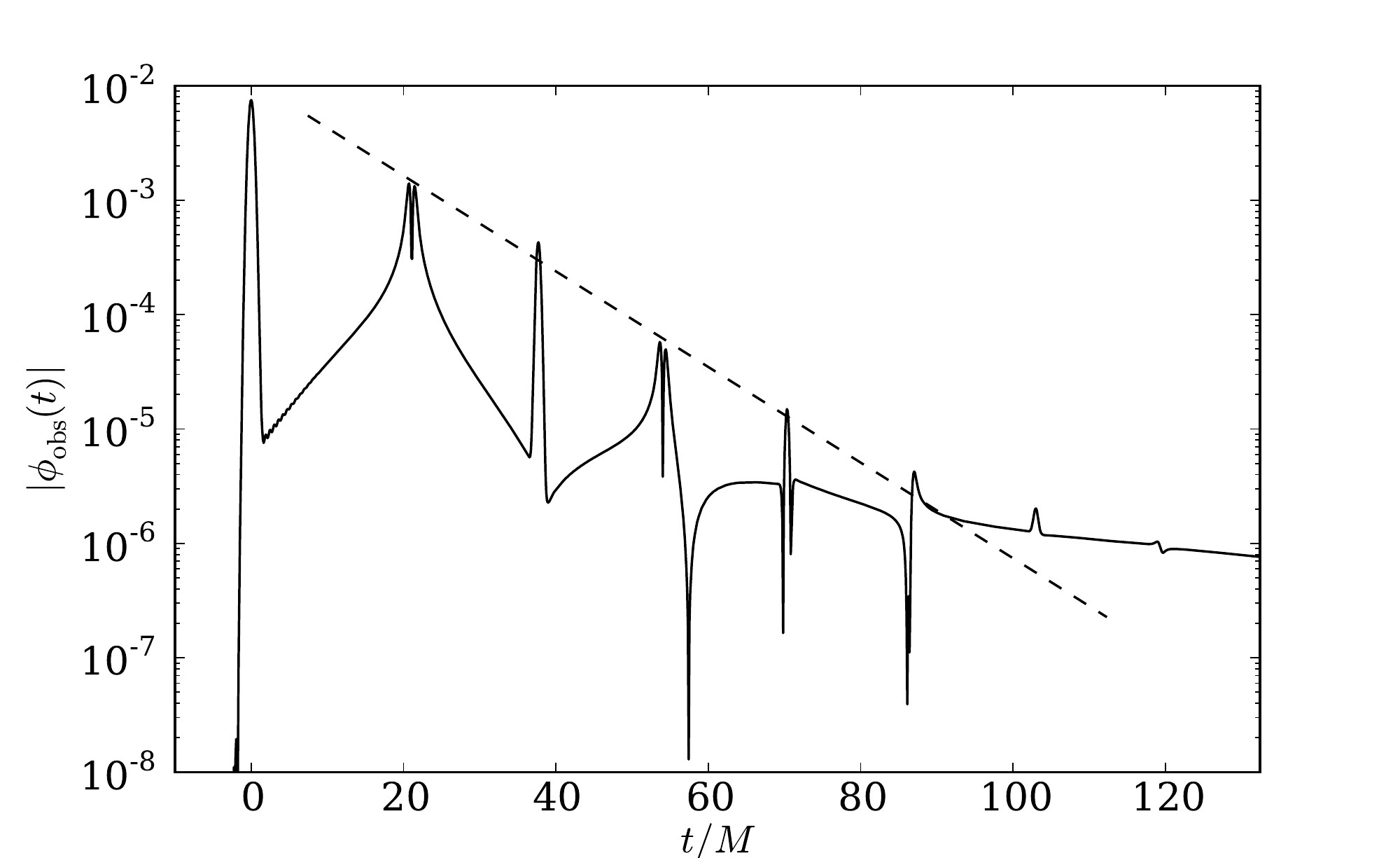}
	\caption{Amplitude decay of the field in
          Fig.~\ref{fig:logplot}. The dashed line shows the
          exponential decay implied by the imaginary part of the $n=0$
          overtone of the quasinormal mode in the large $\ell$ limit
          (\ref{eq:largeLqnm1}), or equivalently the Lyapunov exponent
          $\lambda$ of the null geodesics near the unstable photon
          orbit.}
	\label{fig:ampdecay}
\end{figure}

Figure \ref{fig:logplot} indicates that the amplitude of each caustic
echo decays exponentially with time until the backscatter off the
background curvature dominates. Zooming into the interval before the
tail-dominated domain, Fig.~\ref{fig:ampdecay} shows visual agreement
between the data and the theoretically predicted decay rate of
$\lambda = 1/(2 \sqrt{27}) M^{-1} \approx 0.096225 M^{-1}$.  For
comparison, the decay rate associated with the peak amplitudes of the
even (odd) caustic echoes obtained by fitting to 
$A e^{-\lambda_{\rm num} t}$ gives $\lambda_{\rm num} = 0.103 M^{-1}$ ($0.097M^{-1}$).

The arrival and decay times of caustic echoes approach their predicted
values from the geometrical optics approximation (or equivalently the
large $\ell$ quasinormal mode limit) very fast. The first few echoes
following the direct signal, however, deviate from these limiting values
considerably (Fig.~\ref{fig:period1}). This deviation is due to the
propagation of the first caustic echo along a null geodesic that is
farther away from the photon sphere than the subsequent echoes.

\begin{figure}[ht]
	\includegraphics[width=\columnwidth]{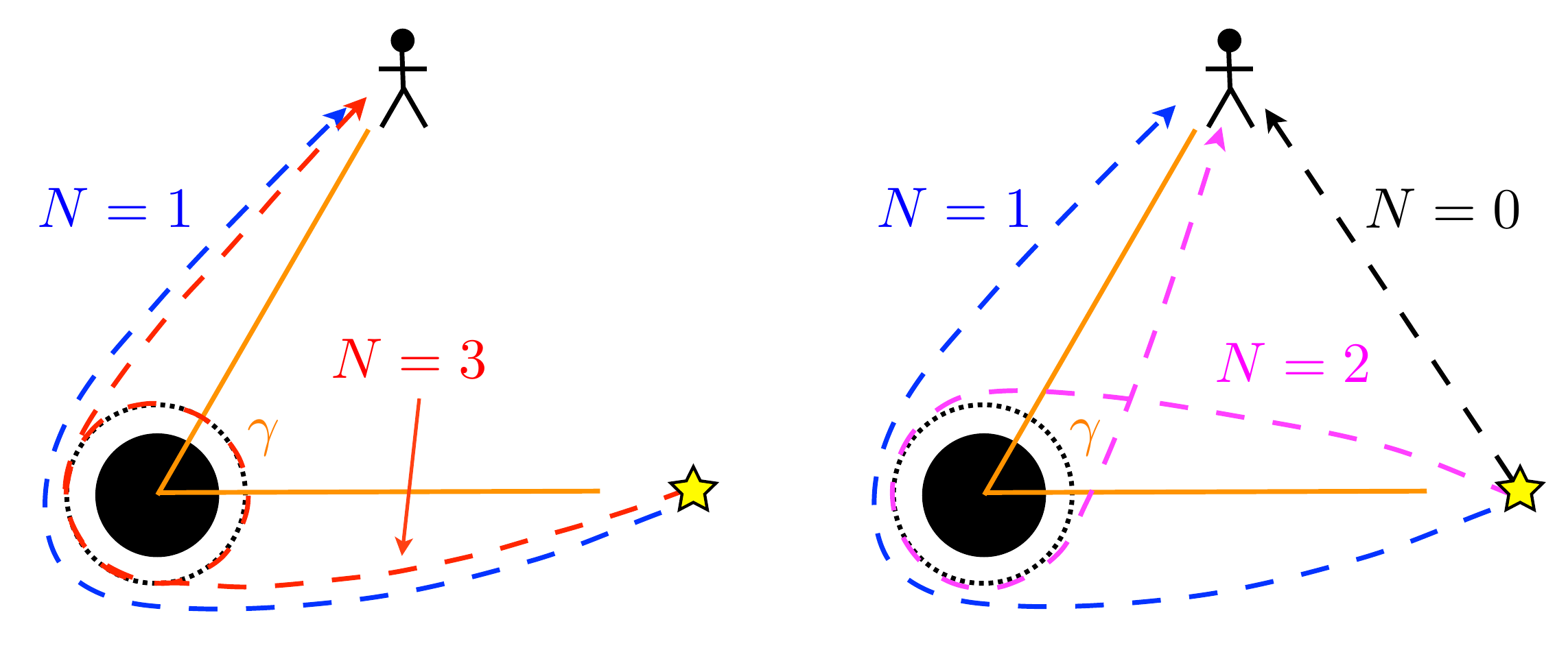}
	\caption{Cartoon depicting several null geodesics connecting
          the source (yellow star) to the observer (stick figure) for
          a generic relative angle, $\gamma \in [0, \pi)$. The black
            hole is represented by a solid black disk and its unstable
            photon orbit by a dotted circle.  Left: The null geodesic
            paths corresponding to the first two odd caustic echoes.
            Right: The null geodesic paths corresponding to the first
            three signals.}
	\label{fig:geodesics1}
\end{figure}

A schematic diagram of the propagation of the first three signals is
given in Fig.~\ref{fig:geodesics1}.  We label the echoes by the order
in time in which they appear, or equivalently by the number of
caustics $N$ that the wavefront has passed through before leaving
the photon sphere.  Consider the the first pulse after the direct
signal that reaches the observer, the $N=1$ echo.  The pulse travels
along the null geodesic connecting the source to the observer wrapping
partially around the far side of the black hole. The angle swept
through by this null geodesic is $\Delta \varphi_1 = 3\pi/2$.  The
null geodesic corresponding to the $N=3$ echo makes a full orbit
around the black hole so that the angle swept through is $\Delta
\varphi_3 = 7 \pi /2$.  For any relative angle between the source and
observer, $\gamma \in [0, \pi]$, it is clear from Fig.~\ref{fig:geodesics1} 
that the time {\it between} successive {\it even or odd} caustic echoes approaches
the period for a null geodesic to make a full orbit around the photon sphere; this 
 is a universal property of the black hole
independent of the source-observer configuration.

For generic source-observer configurations, the time between caustic
echoes depends on the angle $\gamma$ between the source
and observer.  The times between successive caustic echoes are given by
\begin{align}\label{eq:angle}
	\Delta T_N (\gamma) = \left\{ 
	\begin{array}{lcc}
	\displaystyle\frac{ \gamma }{ \pi } \, T_{\rm full} &, & N {\rm ~even} \\
		& & \\
	\displaystyle\frac{ \pi - \gamma }{ \pi }\, T_{\rm full} &, & N {\rm ~odd}
						\end{array}
						\right.
\end{align}
The cases $\gamma=0,\pi$ are degenerate because $\Delta T_N = 0$ for either an even or odd echo. These observers see a
twofold cycle as opposed to a fourfold cycle, and the peak amplitude
of the signal is amplified as seen in Figs.~\ref{fig:3D_1}--\ref{fig:3D_4}.  
The amplification cannot be treated in the
geometrical optics approximation because the amplitude at caustic points is unbounded in
this limit.  We discuss the twofold cycle and the
energy amplification for degenerate observers in
Sec.~\ref{sec:twofold}.

\subsection{Profiles of caustic echoes}
\label{sec:hilbert}

The profiles of caustic echoes transform by propagating through
caustics, which leads to a fourfold cycle
\cite{Ori,Casals:2009xa,Casals:2010zc}.  In this section, we analyze in detail
 the structure of the pulse profiles.

\begin{figure}[ht]
	\includegraphics[width=\columnwidth]{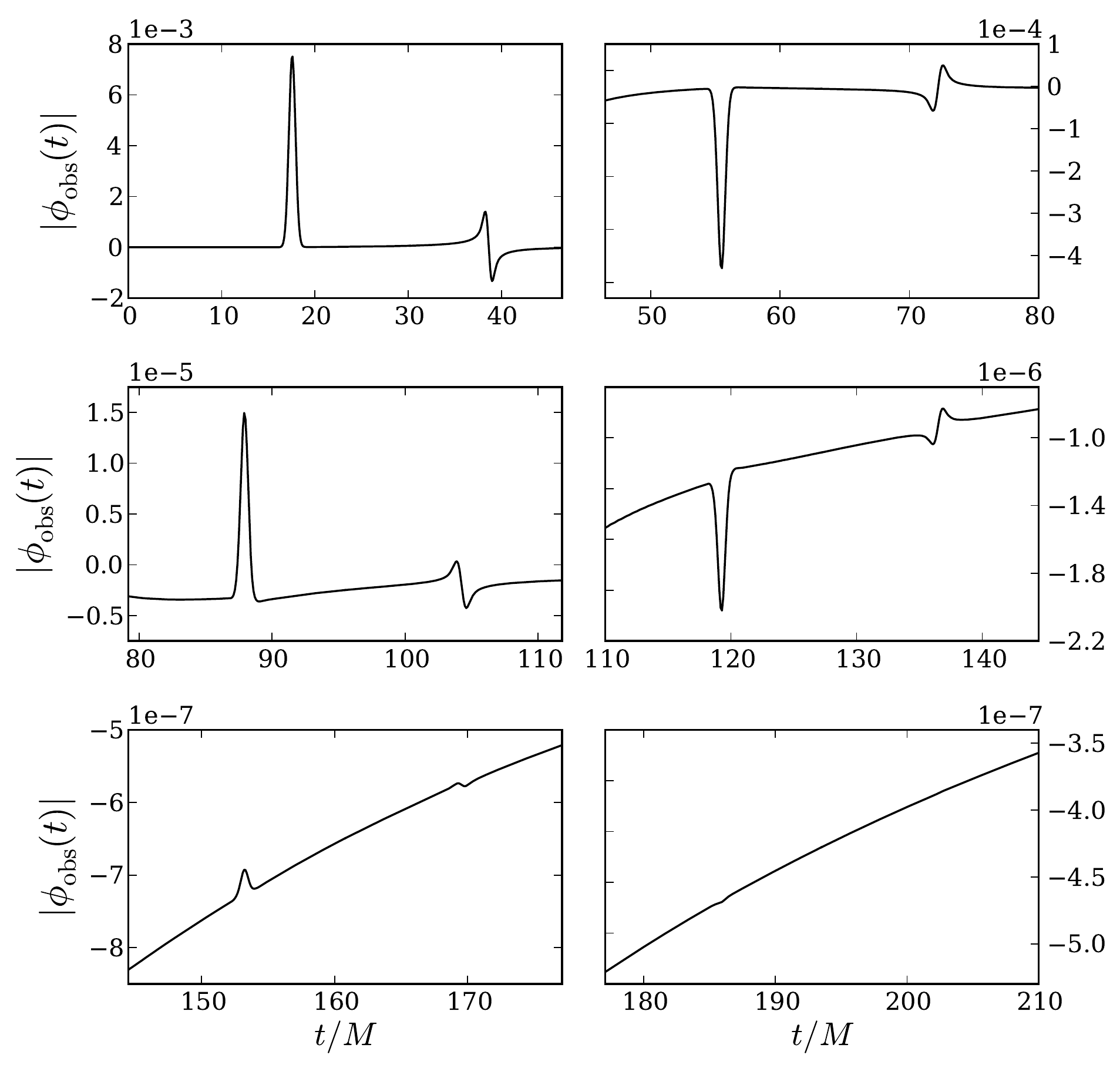}
	\caption{The first 12 pulses (direct signal plus 11
          caustic echoes) seen by an observer on the positive $z$ axis
          at future null infinity. The profiles of the caustic
          echoes are repeated in a fourfold cycle as is clearly seen
          in the first two rows. In the last row, the tail effect from
          backscattering dominates the signal.}
	\label{fig:4fold}
\end{figure}

Figure \ref{fig:4fold} shows the profiles for the same source-observer
configuration as in the previous section. The first pulse in the upper
left panel is the direct signal and the subsequent pulses are the caustic
echoes. The fourfold cycle presented in each row repeats itself until
the signal is dominated by the tail (third row of
Fig.~\ref{fig:4fold}). More caustic echoes are visible in the local
decay rate (Fig.~\ref{fig:tail}).

Using the geometrical optics approximation in Appendix \ref{app:hilbert},
we show that passing through a caustic induces a phase shift of the
wavefront in the frequency domain by $-\pi/2$ radians and is equivalent to a
Hilbert transform (\ref{eq:hilberttransform1}).  The phase change of
light traveling through a caustic is a well-known effect in optics,
called the Gouy phase shift, and was discovered in 1890
\cite{Gouy} (for a more recent discussion see Chapter 9.5 of \cite{Hartmann:2009}). 
The fourfold structure of the Green function is then a
simple consequence of trapped null geodesics passing through caustics
with each passage causing a phase shift by $-\pi/2$, completing a full
cycle of $2\pi$ in four echoes.

The Green function in the geometrical optics approximation is given by
(\ref{eq:Ngreencompact})
\begin{align}
	G( x, x' ) \sim  {} & \sum_{N=0}^\infty \big| A_0 (\bx, \bx') \big| \, \theta ( t- t' - T_{N-1} (\bx, \bx') ) \nonumber \\
		& {\hskip0.25in}  \times {\rm Re} \big\{ H_N \big[ \delta ( t - t' - T_N (\bx, \bx') ) \big]  \big\}. 
\label{eq:greenforshow}
\end{align}
Here, $T_N (\bx, \bx')$ is the coordinate time along a null geodesic
connecting the spatial points $\bx$ and $\bx'$ and passing through $N$
caustics, $H_N$ denotes $N$ applications of the Hilbert transform (thus
phase shifting its argument by $- N \pi / 2$), the step function
ensures causality (with $T_{-1} (\bx, \bx') \equiv 0$), and $A_0 (\bx,
\bx')$ is the leading order field amplitude in the geometrical optics
limit (see Appendix \ref{app:hilbert} for the derivation).  The field is
obtained by the convolution of the Green function with the source
$S(x)$
\begin{align}
	\label{eq:convol}
	\phi (x) & = \int _{x'} G (x, x') S (x')  \sim \sum_{N=0}^\infty \phi_N (x),
\end{align}
where $\int_{x'} \equiv \int d^4 x' \sqrt{-g(x')}$ with $g$ the
determinant of the spacetime metric, and $\phi_N(x)$ is the field
evaluated in the geometrical optics approximation associated with the
passage of the wavefront through $N$ caustics.

Equation \eqref{eq:convol} gives us a quantitative prescription for
the profiles of the wavefronts after propagation through $N$
caustics. For $N=0$, the Hilbert transform $H_0$ returns its
argument. Hence the direct signal obtained from the convolution of the
approximate Green function (\ref{eq:greenforshow}) with the Gaussian
source \eqref{eq:source} is proportional to a Gaussian as confirmed in
the upper left panel of Fig.~\ref{fig:4fold}. For $N=1$ the Hilbert
transform $H_1$ applies a phase shift by $-\pi/2$
radians to the signal entering the caustic. For the delta distribution in (\ref{eq:greenforshow}) this is
given by (\ref{eq:oneoversigma1}). Interestingly, (\ref{eq:oneoversigma1}) has a different singularity structure than the delta distribution and has support in the interior of the forward light cone of $\bx'$. Hence, this contribution adds to any tail piece from backscattering once the wavefront has passed through an odd caustic. The caustic thus smears out the (sharp) delta distribution that gets refocused to a delta distribution when passing through the next caustic.

For the first caustic echo we get from
\eqref{eq:convol}
\begin{align}
	\phi_1(x) = {} & \frac{ 1 }{ (2\pi \sigma^2 ) ^2 } \int _{\bx'}  | A_0 (\bx, \bx') | \exp \left(- \frac{(\bx' - \bx_0)^2}{2 \sigma^2} \right) \nonumber \\
		\times & {\rm Re} \bigg[ \int_{-\infty}^\infty \frac{ d\omega }{ 2\pi }  \, \exp \left( - \frac{ i \pi }{2} - i \omega \big( t - T_1(\bx, \bx') \big)  \right)  \nonumber \\
		&  {\hskip0.4in} \times \int_{-\infty}^{t - T_1 (\bx, \bx')} {\hskip-0.15in} dt' \exp \! \left(i \omega t'  - \frac{ (t' - t_0)^2 }{2\sigma^2} \right) \!\! \bigg].
\label{eq:N1convolution1}
\end{align}
Integrating over $t'$ (assuming $t - T_1 (\bx, \bx')  \gtrsim t_0 + \sigma$ so that the upper integration limit is replaced by $\infty$) and $\omega$, and assuming that the spatial part
of the Gaussian is sufficiently narrow to be replaced by a delta
distribution yields
\begin{align}
	\phi_1(x) \approx - | A_0 (\bx, \bx_0 ) |  D \big( t - t_0 - T_1 (\bx, \bx_0), \sigma \big),
\label{eq:phi1b}
\end{align}
where 
\begin{align}
	D( y, \sigma) \equiv e^{ - y^2/ (2\sigma^2) } {\rm erfi} \left( - \frac{ y }{ \sqrt{ 2 } \, \sigma } \right)
\label{eq:dawsonian1}
\end{align}
is related to Dawson's integral \cite{Duoandikoetxea}. 
We will refer to (\ref{eq:dawsonian1}) as the ``Dawsonian'' below.

\begin{figure}[ht]
	\includegraphics[width=\columnwidth]{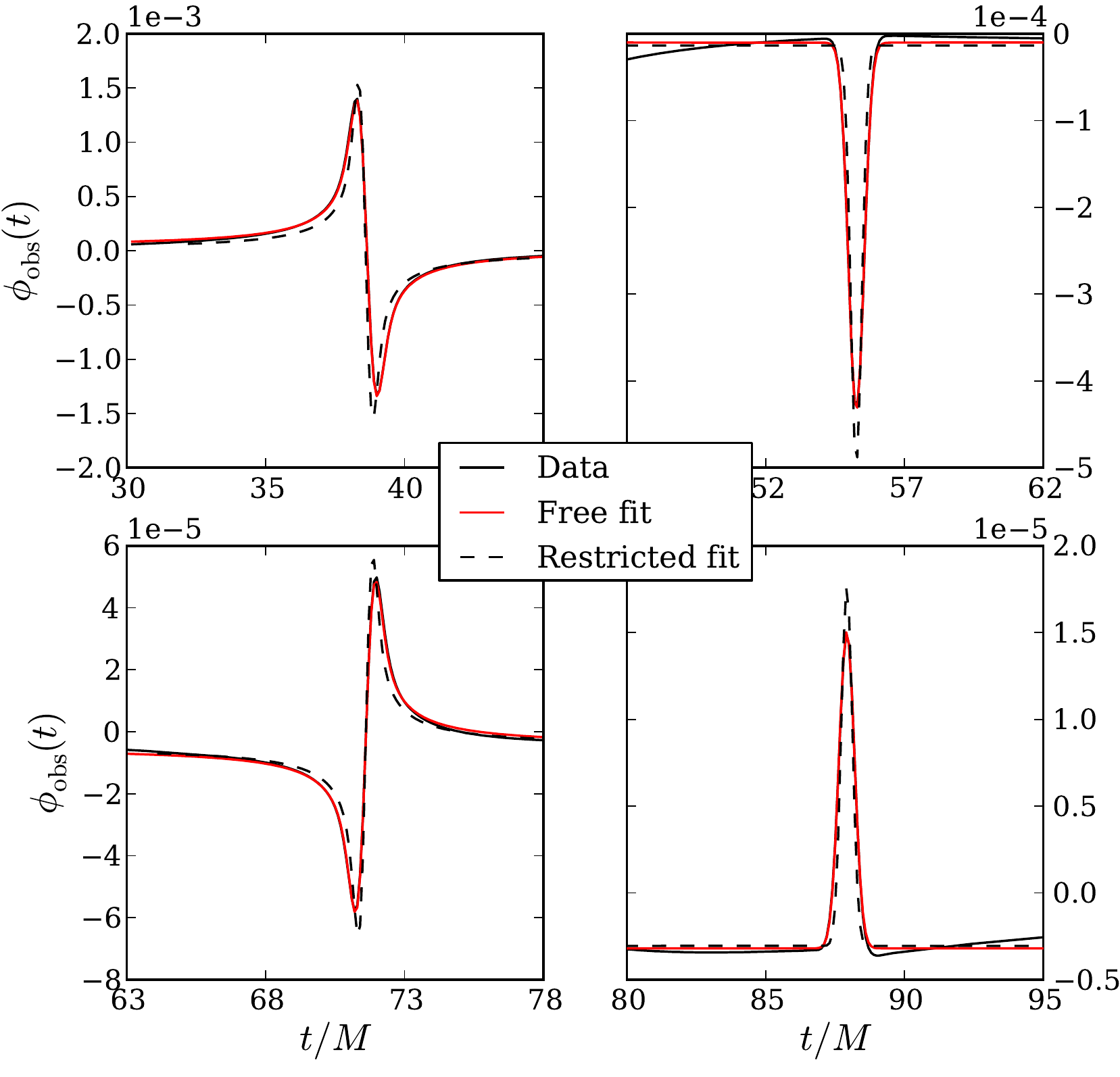}
	\caption{The first four caustic echoes in the numerical
          solution (black) along with two fits to a Dawsonian using the geometrical
          optics approximation (red and dashed black). The numerical
          solution is fit to Eq.~(\ref{eq:phi1b}) with the amplitude,
          time of arrival, and width as fitting parameters. The fit is so well that the red and the black lines are hardly distinguishable. The black
          dashed line is a restricted fit where only the amplitude is
          fitted for and the time of arrival and the width are set by
          the numerical solution and the Gaussian source width,
          respectively. The quality of the fits 
          indicate that the geometrical optics approximation captures
          the profiles of the caustic echoes accurately via the
          Hilbert transform.}
	\label{fig:dawson}
\end{figure}

The upper left panel of Fig.~\ref{fig:dawson} shows the first caustic echo (e.g., the second
pulse in the upper left panel of Fig.~\ref{fig:4fold}) together with
two different fits to a Dawsonian using (\ref{eq:phi1b}).  The black line is the
numerical data and the red line is a fit to $a D( t - b, c) + d$ with
$a = - 2.8 \times 10^{-3}$, $b = 38.64M$, $c = 0.28M$, and $d = 2.4
\times 10^{-5}$. Note that the best-fit value for $c=0.28M$ differs
substantially from the original width of the Gaussian source of
$\sigma = 0.2M$.  This free fit is remarkably good over a large interval
around $T_1 (\bx, \bx_0) = b$.

The black dashed line in Fig.~\ref{fig:dawson} shows the approximation
(\ref{eq:phi1b}) with $\sigma = 0.2M$ (i.e., the width of the Gaussian
source) and $T_1 (\bx, \bx_0)$ fixed to the value extracted from the
numerical simulation; only the amplitude is fitted for. Despite the
crudity of the approximations, the restricted fit captures qualitative
features of the $N=1$ echo rather well. The agreement indicates that
the Green function in the geometrical optics approximation is suitable
to describe the propagation of wavefronts through caustics in terms of
Hilbert transformations. The remaining panels in Fig.~\ref{fig:dawson}
for the next three caustic echoes show similar results.

Our discussion so far covers wavefront propagation through caustics
for relatively narrow pulses providing an approximation to the Green
function. We confirmed the action of a Hilbert transform at caustics
using other waveforms as well.

\begin{figure}[ht]
	\includegraphics[width=\columnwidth]{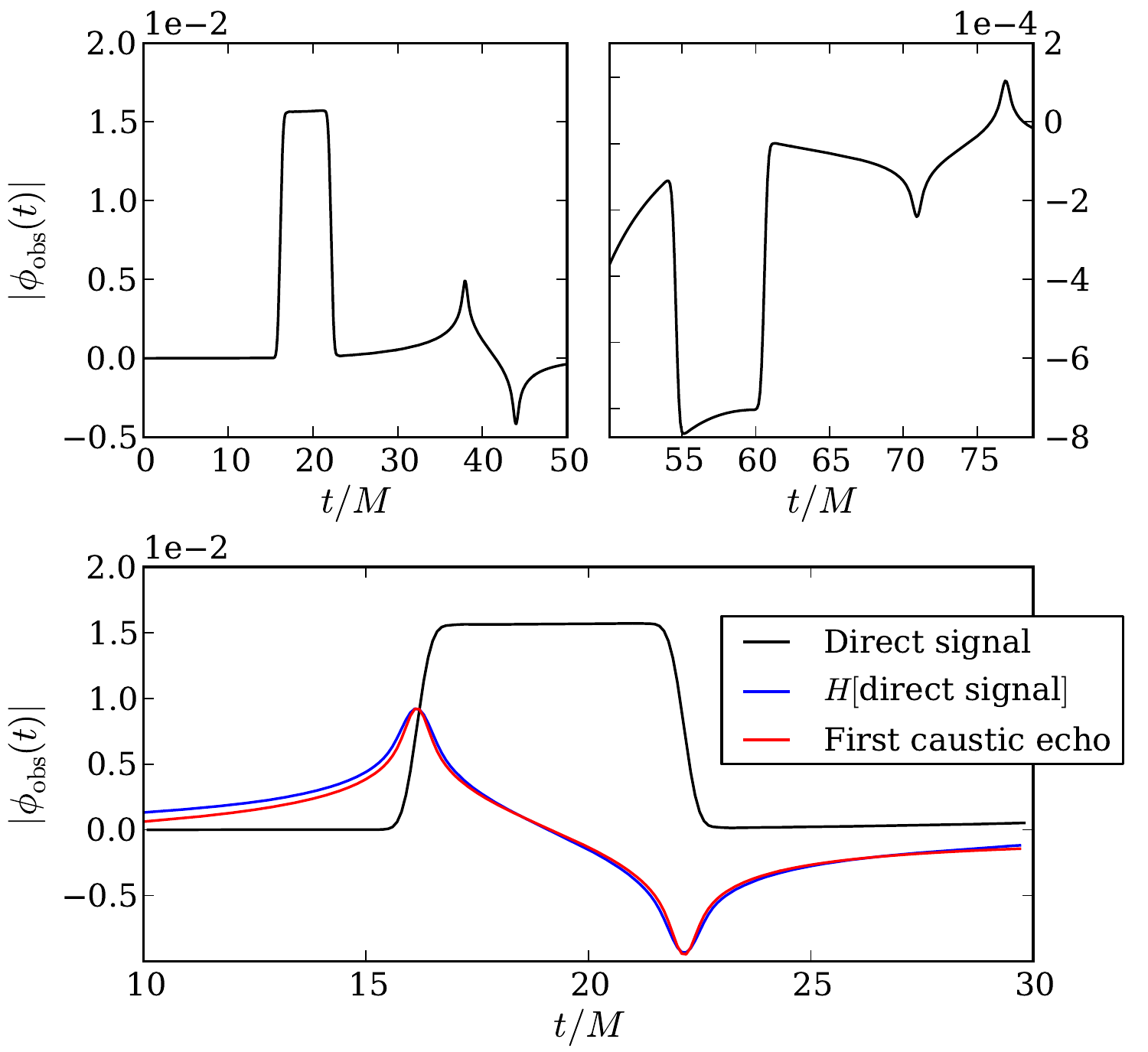}
\caption{{\bf Top}: The caustic echoes of a square pulse source
  lasting for $6M$ and located at $6M$ on the positive $x$ axis as
  seen by an observer on the positive $z$ axis at future null
  infinity. {\bf Bottom}: The direct signal of the square pulse
  (black), its Hilbert transform (blue), and the first caustic echo
  appropriately shifted and rescaled to match the amplitude of the
  blue curve. Despite the long time scale of the square pulse, its
  corresponding wavefronts are still Hilbert transformed through
  caustics.}
\label{fig:hilbert}
\end{figure}

The top panels in Fig.~\ref{fig:hilbert} show a fourfold cycle
triggered by an initial square profile, that is, a pulse
localized in space but persisting for a longer time scale than the
time scale of the black hole.  In the lower panel of
Fig.~\ref{fig:hilbert} we plot the direct signal (black curve) and
compare its Hilbert transform (blue curve) to the rescaled and shifted
 profile of the $N=1$ caustic echo from the numerical simulation
(red curve). The agreement between the blue and the red curves implies
that the wavefronts are indeed Hilbert transformed through caustics.

\subsection{The twofold cycle and the caustic line}
\label{sec:twofold}

Equation \eqref{eq:angle} indicates that for angles $\gamma=0,\pi$
between the source and the observer, the structure of the measured
signal is degenerate.  This is the caustic line where caustic echoes
merge (the $x$ axis in our simulations). Even and odd echoes arrive at
the same time at the caustic line, therefore we obtain a {\it
  twofold} cycle as opposed to a fourfold one. Hence, the number of echoes in a cycle is observer-dependent but the existence of a cyclic feature is universal and associated with the trapping of energy near the black hole.

\begin{figure}[ht]
	\includegraphics[width=\columnwidth]{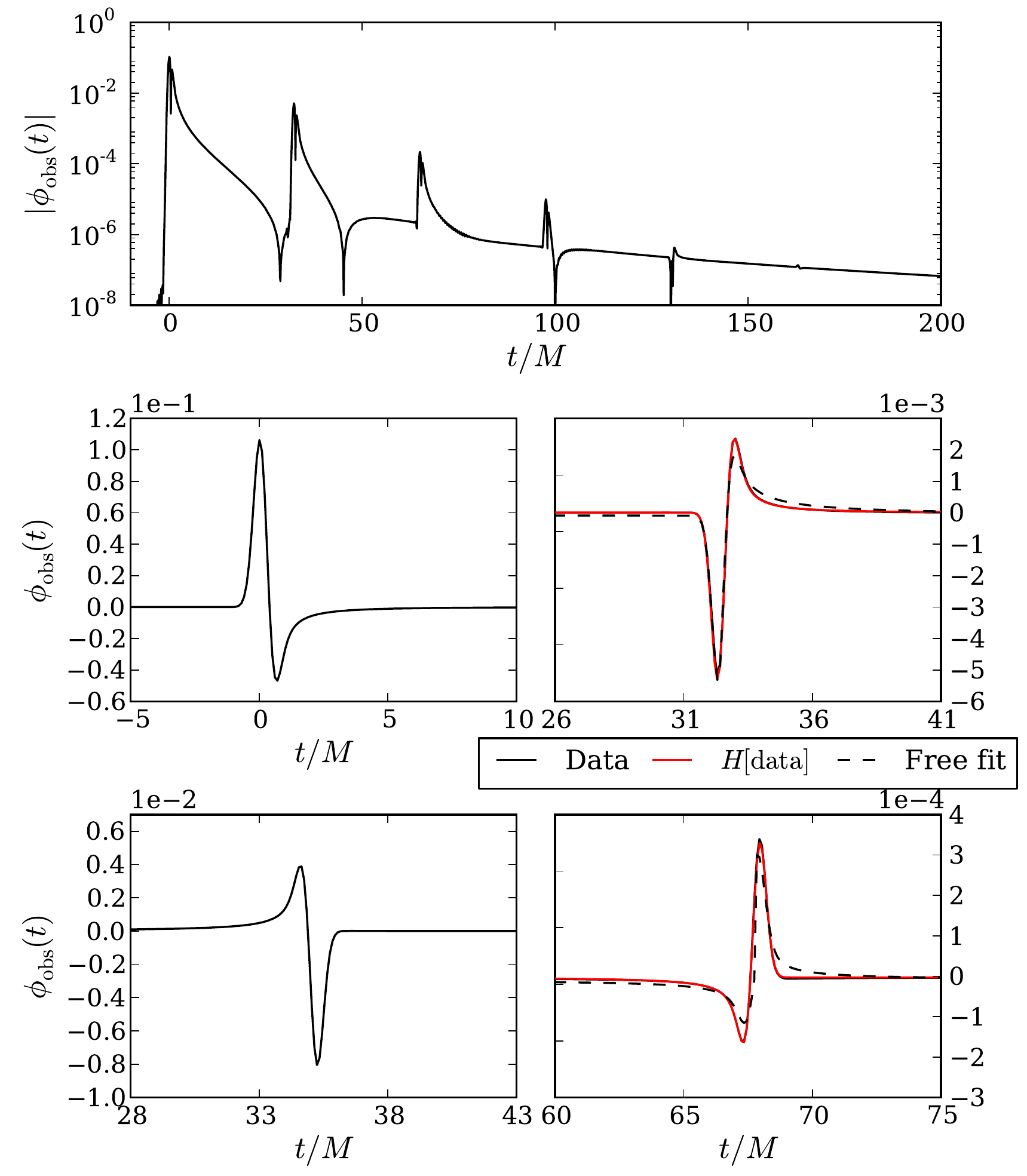}
	\caption{{\bf Top:} The evolution of the scalar field as
          measured by an observer at future null infinity on the
          negative $x$ axis. {\bf Middle:} Degenerate echoes as
          measured by the same observer. Because the source is also on
          the $x$ axis, the observer only sees a twofold cycle. Also
          shown is the Hilbert transform of the signal (red) in the
          left panel (rescaled and shifted to compare with the caustic
          echo) and the free fit to the geometrical optics
          approximation (dashed). {\bf Bottom:} Same as above except
          that the observer is on the positive $x$ axis. Note the
          differences in the caustic echo profiles.}
	\label{fig:2fold}
\end{figure}

Consider the top panel of Fig.~\ref{fig:2fold}. The first pulse is
formed by the direct signal and the $N=1$ caustic echo. The next pulse
arrives after a full revolution time $T_{\rm full}$ and consists of the $N=2, 3$
caustic echoes. (Compare Fig.~\ref{fig:geodesics1} or
Figs.~\ref{fig:3D_1}-\ref{fig:3D_4} with an observer along the caustic
line to visualize the simultaneous arrival of echoes.)

The echoes on the caustic line also undergo a Hilbert transformation.
The middle panel of Fig.~\ref{fig:2fold} shows the first ($N=0,1$) 
and the second ($N=2,3$) degenerate echoes for an
observer on the negative $x$ axis ($\gamma = \pi$), while the bottom
panel of Fig.~\ref{fig:2fold} shows the first ($N=1,2$) and the
second ($N=3,4$) degenerate echoes for an observer on the
positive $x$ axis ($\gamma = 0$). The $N=0,1$ degeneracy for the
$\gamma = \pi$ observer leads with a Gaussian profile which is Hilbert
transformed at its peak to a Dawsonian for the trailing half of the
signal. The $N=1,2$ degeneracy for the $\gamma = 0$ observer leads
with a Dawsonian profile which is Hilbert transformed at its zero
crossing to a Gaussian for the trailing half of the signal.  The next
degenerate echoes are the Hilbert transforms of the previous ones as
indicated in the middle and bottom right panels of
Fig.~\ref{fig:2fold}. The red curves are the Hilbert transforms of the
previous degenerate echoes to the left, suitably shifted in time and
rescaled in amplitude for comparison.

The amplitude of the wave becomes singular along the caustic line in
the geometrical optics approximation. However, given that the echoes
can be characterized via Hilbert transforms whenever the wavefront
passes through a caustic, the time dependence of the signal may be
captured by the geometrical optics approximation. To test this, we fit
the data to the profile implied by the convolution of the Green function
in the geometrical optics approximation (\ref{eq:greenforshow}) with the source (\ref{eq:source}).
More specifically, the $N=0,1$ degeneracy for the $\gamma=\pi$ observer is fit to
\begin{align}
	a \, e^{-(t-b)^2/(2 d^2)} + c \, \theta ( t - b) D( t- b, d) 
\end{align}
where $b$ is approximately the time of arrival $T_0 = T_1$, while for the $N=1,2$
degeneracy for the $\gamma = 0$ observer the fit is
\begin{align}
	a \, D (t-b, d) +  c \, \theta(t-b) \, e^{-(t-b)^2/(2 d^2)} 
\end{align}
with $b$ approximately $T_1 = T_2$. In both cases, 
the amplitudes ($a,c$), the time of arrival ($b$), and the width are arbitrary ($d$). The results of this free
fit are shown in the middle and bottom right panels of
Fig.~\ref{fig:2fold} (dashed lines). Despite its breakdown, the
geometrical optics approximation captures the basic time-dependence of
the profiles fairly well. 

The wavefront is focused and its amplitude is magnified along the
caustic line (see Figs.~\ref{fig:3D_1}-\ref{fig:3D_4}). We next
characterize this magnification at the caustic relative to the echoes
it generates in relation to the scale of the wavefront. The
computations of such relations go beyond the geometrical optics
approximation and require elements of geometric diffraction theory. In
\cite{KayKeller} Kay and Keller show that the peak amplitude at a
caustic with a perfect focus increases as $\sigma^{-1/2}$, where
$\sigma$ provides the scale for the width of the wavefront.

Note that the caustic and its echoes have different shapes. Therefore,
a comparison of their peak amplitude might be misleading. Instead, we
compare the energy radiated to an observer at infinity during a certain time interval $I$,
\begin{align}
	E= \frac{1}{2} \int_I dt \, \big| \dot{\phi} (t) \big|^2 .
\end{align}

\begin{figure}[ht]
	\includegraphics[width=\columnwidth]{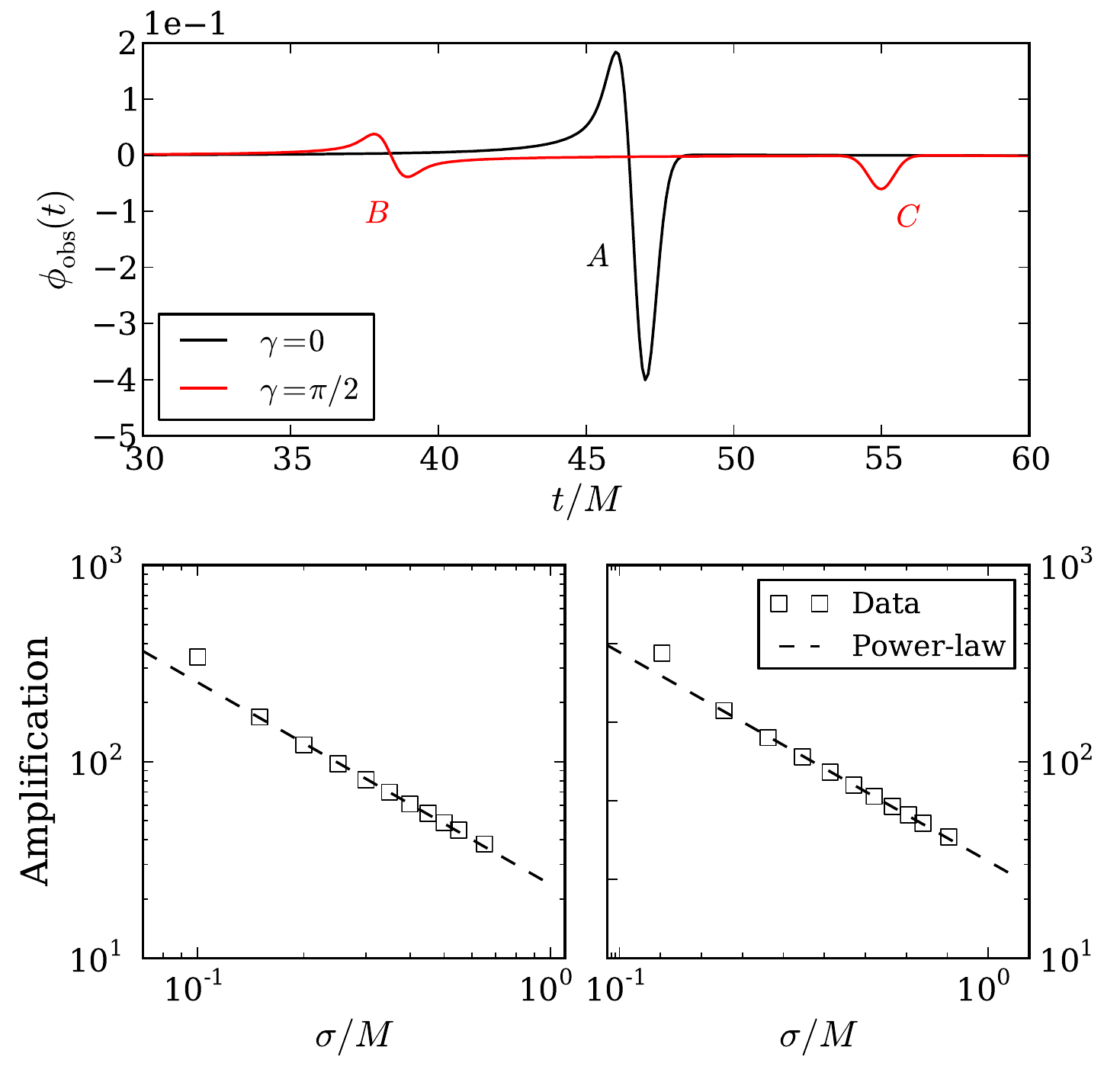}
	\caption{{\bf Top}: A caustic ($A$) at $\gamma=0$ (black), 
	the $N=1$ echo ($B$),
          and the $N=2$ ($C$) echo along $\gamma=\pi/2$ (red). {\bf
            Bottom}: The amplification of the energy of $A$ relative
          to $B$ (left) and $C$ (right) along with an inverse power law fit (dashed).}
	\label{fig:amplification}
\end{figure}

To make sure that the amplification factor is not influenced by the
shape of the echoes, we compute it with respect to both Dawsonian and
Gaussian shapes. The compared signals are plotted in the top panel of
Fig.~\ref{fig:amplification}. We label the measurement of the caustic
($N=1,2$) by an observer at $\gamma=0$ as $A$. An observer at
$\gamma = \pi/2$ measures not the caustic but its echoes. The
Dawsonian echo ($N=1$) is labeled as $B$, the Gaussian one ($N=2$) 
as $C$.

We compute the energy integral over a sufficiently large time interval
centered at the pulse so that the energy difference to a larger
interval is negligible (less than $1\%$). The amplification is then
computed via
\begin{align}
	{\rm Amplification} \equiv \frac{ E_{A} }{ E_{B,C}} .
\end{align}

We plot the amplification on a log-log scale as a function of Gaussian
widths $\sigma$ in the bottom two panels of
Fig.~\ref{fig:amplification}. The amplification factor can be fit
accurately by an inverse power law of the form $a \sigma^p$ where
$(a_B, p_B) = (23.7, -1.03)$ (bottom left) and $(a_C, p_C) =
(26.1,-1.02)$ (bottom right). Hence, the data suggests that the energy
amplification at the caustic goes as $\sigma^{-1}$ as the width of the
Gaussian source becomes more narrow. This seems in accordance with the calculation of Kay and Keller for caustics with a perfect focus 
\cite{KayKeller}.

We emphasize, however, that we cover only a small range of scales 
($\sigma \in [0.1,0.6]$) and that we did not mathematically derive the
scale dependence of the energy at caustics formed by a Schwarzschild
black hole. Including smaller values for $\sigma$ will most likely follow an inverse power law but may change the value in the fit. Therefore, our numerical result should be interpreted carefully. 

There are extensive studies on the detailed behavior of
solutions to wave equations near caustics, including not only their
energy magnification but also their shapes in different directions
away from the caustic line. Such an analysis is outside the scope of
this paper but provides an interesting subject for future research.

\subsection{Propagation within the light cone}
\label{sec:tail}

We have shown that essential properties of the Green function can
already be understood in the geometrical optics limit. We also
discussed effects due to the finite wavelength of the wave, which
plays a role in the energy magnification at caustics.  In this
section, we consider features at the largest scale, the spacetime
curvature.

\begin{figure}[ht]
\includegraphics[width=0.95\columnwidth]{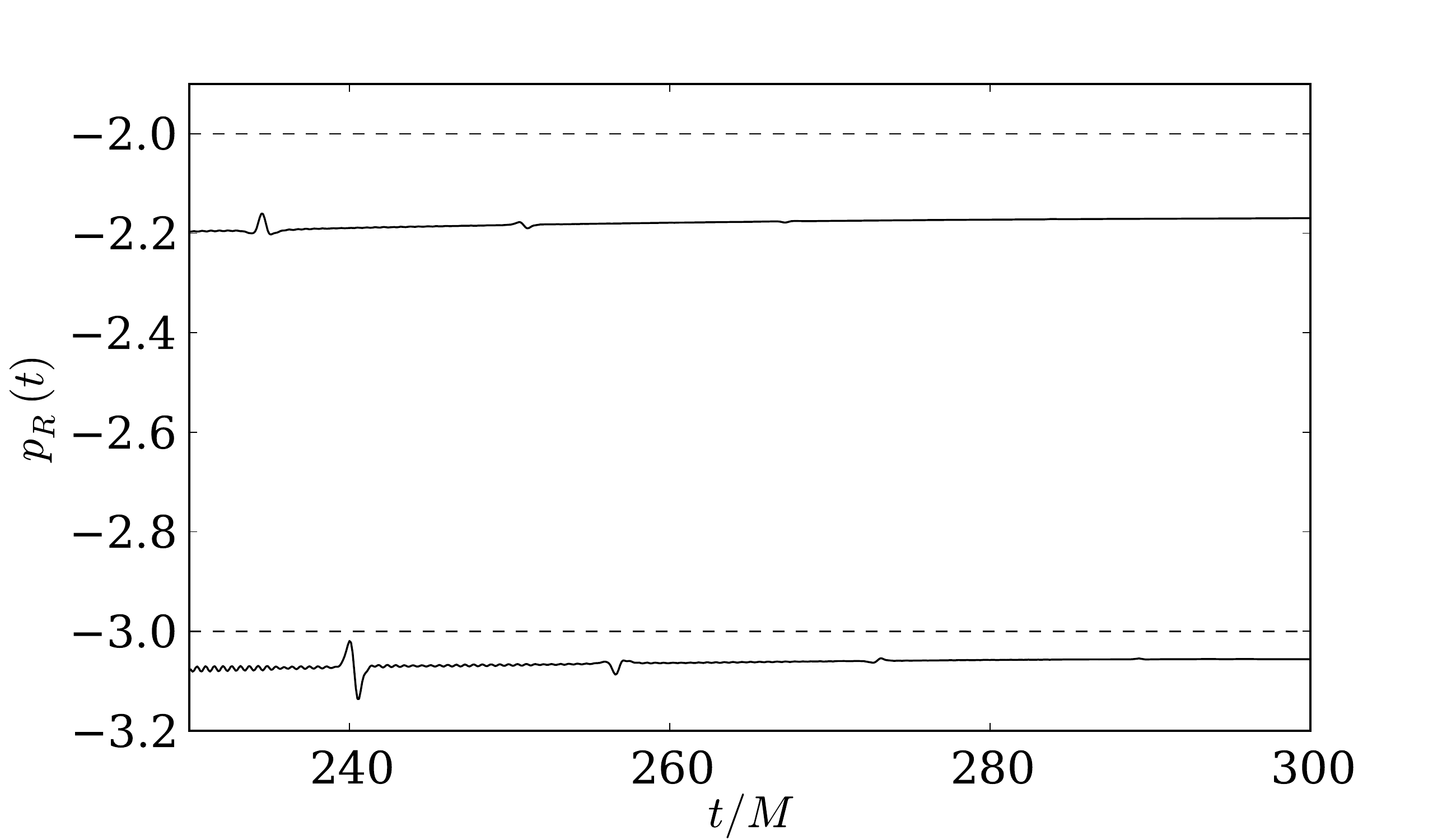}
\caption{Polynomial decay rates asymptotically approaching $p=-2$ for
  an idealized observer at null infinity (top solid curve), and $p=-3$
  for a finite distance observer at $r=4M$ (bottom solid curve). The
  theoretical asymptotic decay rates are plotted as dashed
  lines. Caustic echoes are still visible in the local power plot at
  late times.
\label{fig:tail}}
\end{figure}

It is well-known that the Green function propagates also inside the
null cone in curved spacetimes due to the backscatter of the field off the
background curvature.  The field within the null cone decays
polynomially following the so-called Price power law \cite{Price:1971fb}.
The rate of the decay for idealized observers at null infinity is
different than the rate at finite distances \cite{Gundlach:1993tp}.
While the null infinity rate is the relevant one with respect to
idealized observers, the finite distance rate is the one that effects
the self-force on a particle. Both rates are plotted
in Fig.~\ref{fig:tail}.

The theoretical decay rates are valid asymptotically in time. One can
obtain a good estimate of the rates by computing a local in time rate
along the worldline of an observer at $r=R$ defined by
\be\label{eq:rate} p_{R}(t) = \frac{d\ln |\phi(t,R)|}{d\ln t}
. \ee The theoretical asymptotic decay rate for a generic scalar
perturbation in Schwarzschild spacetime is for idealized observers at
null infinity $p_\infty=-2$ and for finite distance observers
$p_R=-3$. Figure \ref{fig:tail} confirms the approach of the local
power to the asymptotic rate. The theoretical predictions are plotted
as dashed lines.

\begin{figure}[ht]
\includegraphics[width=0.9\columnwidth]{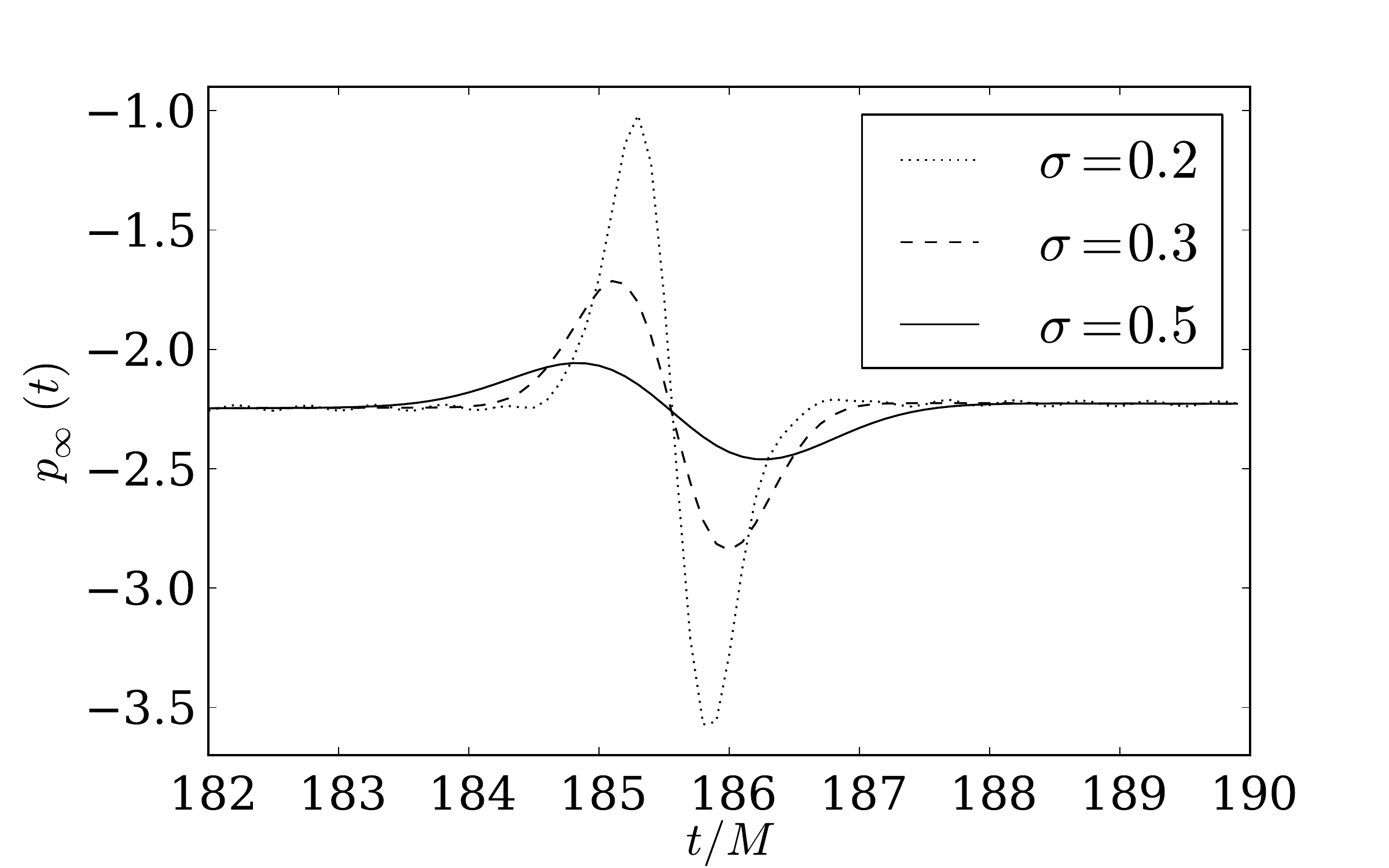}
\caption{The visibility of a caustic echo in the local power at null
  infinity for three different values of the width of the initial
  Gaussian source \eqref{eq:source}. More caustic echoes are visible
  at late times for narrower Gaussians.
\label{fig:narrow}}
\end{figure}

In discussions of gravitational lensing it is often stated that, under
certain assumptions and with ideal instruments, one would see
infinitely many images of a source with an eternal worldline
\cite{Perlick:2010zh}. Considering the geometrical optics limit, one
might be inclined to think that for arbitrarily narrow Gaussians we
would observe arbitrarily many caustic echoes. We show in 
Fig.~\ref{fig:narrow} a caustic echo visible in the local power of the
signal at null infinity for three different values of $\sigma$ for the
Gaussian source \eqref{eq:source}. The caustic echo is more pronounced
for smaller values of $\sigma$ indicating that the narrower our
source, the more caustic echoes we see in the late time signal.

However, it is also clear that for any positive value of $\sigma$ we
can only see a finite number of caustic echoes because the amplitude
of each echo decays exponentially in time whereas the backscatter off
curvature decays only polynomially.  This observation implies that,
even with ideal instruments, we can measure only a finite number of
``images" for any given source because eventually the polynomial decay
will win over the exponentially weakening echo. Proofs of infinitely
many images of sources typically consider the geodesic structure of
spacetimes and neglect backscatter \cite{Hasse:2001by}.

\subsection{Putting the pieces together}
\label{sec:full}

We presented in previous sections a quantitative understanding of each
feature in the evolution of a scalar perturbation triggered by a
narrow Gaussian wave package.  Dissecting the signal observed at
infinity (Fig.~\ref{fig:logplot}) we explained the arrival times, the
exponential decay, the shapes of the echoes, and the late time
behavior of the scalar field. In this section, we bring these elements
together into a heuristic formula that captures the essential features
of the evolution remarkably well.

The shape, the arrival times, and the exponential decay of caustic echoes 
can be accurately described in the geometrical optics limit as
discussed in Secs.~\ref{sec:echoes} and \ref{sec:hilbert}. We
derived an analytic expression for the Schwarzschild Green function in this limit
(Appendix \ref{app:optics}), presented for the first time in this paper as far as we are aware. Using this analytic knowledge we approximate
the signal by
\begin{align} 
	\phi_{\rm geom} & =  A_0 \bigg[ e^{- (t -T_0  )^2 / (2 \sigma_a^2 ) }  + e^{ - \lambda (t  -T_0  ) } \nonumber \\
	& \times {\rm Re} \sum_{n=1}^{N} \theta( t - T_{n-1} - \delta T) \, e^{-  i n \pi / 2  } \nonumber \\
	&  \times \bigg( e^{-  (t - T_n - \delta T)^2 / (2 \sigma_a^2)  }  - i  D ( t - T_n - \delta T , \sigma_a ) \bigg) \bigg] . 
\label{eq:phigeom1}
\end{align}
where $\lambda \approx 0.096225$ is the Lyapunov exponent for unstable
photon orbits, $T_0$ is the time of arrival of the direct signal to
the observer, and $T_{n}$ is the time of arrival for the $n^{\rm th}$
caustic echo (with $T_{-1} \equiv 0$). The latter is computed from
$T_{n-1} + \delta T + \Delta T_n (\gamma)$ using (\ref{eq:angle}). The quantity $\delta T$ accounts for the fact that the time between successive caustic echoes is not immediately given by $\Delta T_n (\gamma)$ (Fig.~\ref{fig:period1}). We found it sufficient to approximate $\delta T$ by $T_2 - T_0 - T_{\rm full}$, which is independent of $\gamma$. Only $T_0$ and $\delta T$ (or $T_2$) are extracted from the numerical simulation.  We
set $A_0=0.016$, $\sigma_a=0.28$, and the number of caustic echoes as
$N=15$ specifically for the observer of Fig.~\ref{fig:logplot}.

\begin{figure}[ht]
	\includegraphics[width=\columnwidth]{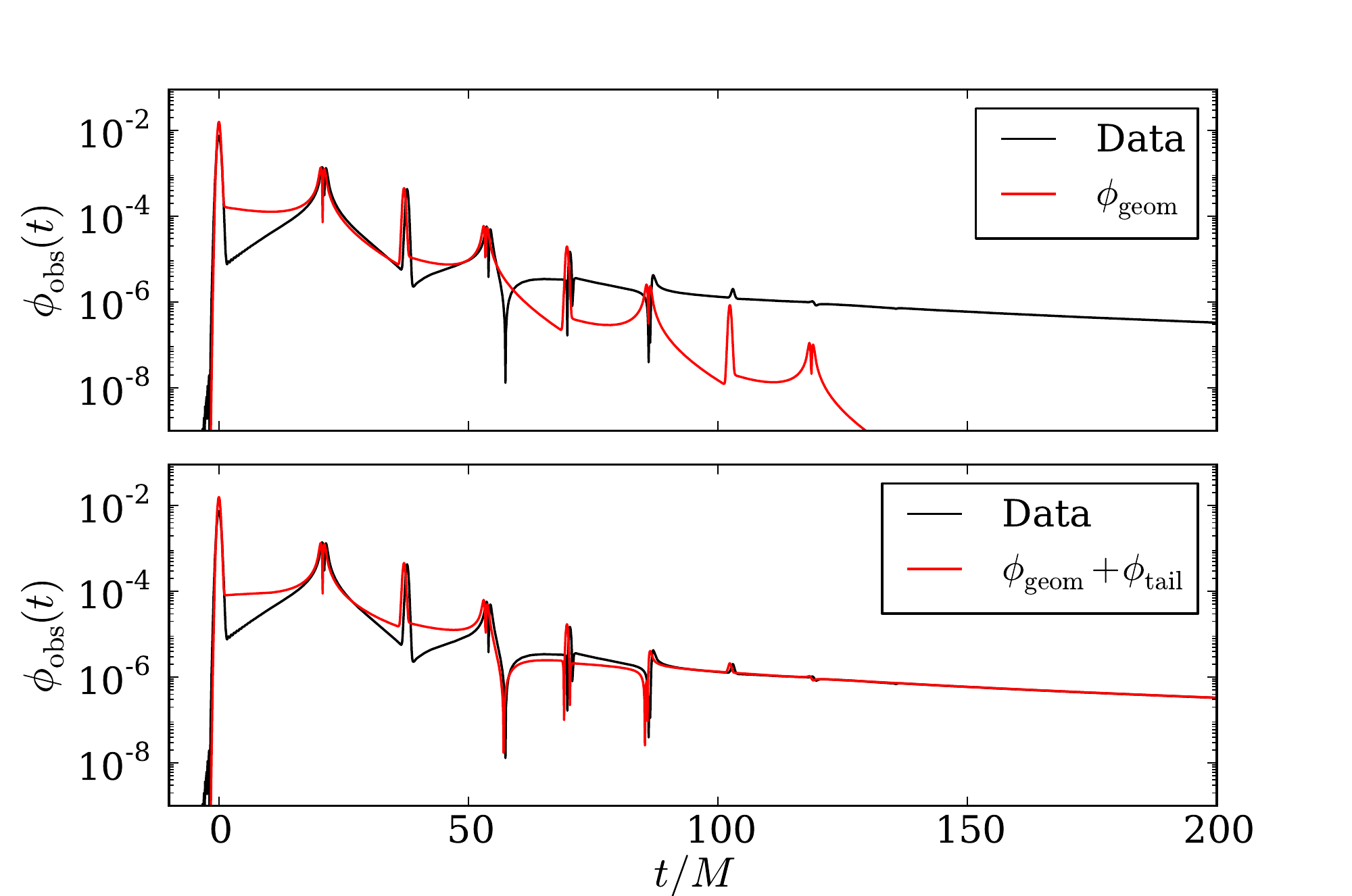}
	\caption{Reconstructing the numerical solution (black) using
          heuristic formulae (red). The top panel shows the structure
          of the field from the geometrical optics limit. The bottom
          panel includes the power law tail contribution. Despite the
          simple nature of the approximations, the field's rich
          dynamics are captured remarkably well.}
	\label{fig:enchilada}
\end{figure}

The geometrical optics formula for the scalar field agrees with the
numerical data in the early part of the signal (top panel of
Fig.~\ref{fig:enchilada}). To match the late time behavior of the
field we use the polynomial decay discussed in Sec.~\ref{sec:tail}.
We include the tail only after the direct signal has reached the observer 
and set 
\begin{align}
	\phi_{\rm tail} = C t^p \, \theta ( t - T_0) ,
\label{eq:phitail1}
\end{align}
with $C= -0.06$ and $p=-2.25$. 
Note that the asymptotic decay rate would 
be $p=-2$. The difference is due to the high order correction terms which
are neglected in the above formula. One can include such correction terms
by constructing more general fitting functions (see for example 
\cite{Bizon:2009wp}). Recently, Casals and Ottewill computed the first three orders in the power law for the Schwarzschild Green function \cite{Casals:2012tb}, which can also be used to improve the description of the tail. We found the simple expression above sufficiently accurate for our purposes. Better expressions for the tail fit might be needed for comparisons with more accurate approximations to the Green function, or for self-force computations. 

Our heuristic formula for the full field
is the sum of the geometrical optics contribution \eqref{eq:phigeom1}
and the tail contribution \eqref{eq:phitail1}. Considering the
simplicity of the assumptions that go into our heuristic expression,
its agreement with numerical data is remarkable (bottom panel of
Fig.~\ref{fig:enchilada}).

Note that the addition of the tail not only improves the agreement
with the data at late times but also captures the zero crossings
between the caustic echoes at $t/M \approx 57$ and $86$. In typical
numerical studies of the tail, the polynomially decaying part of the
signal plays a role only at late times. Here, we see that it improves
the fit between the echoes at relatively early times. This is because
backscattering contributes even right after the direct signal has
passed the observer. This feature is well-known and is captured in 
Hadamard's ansatz for the Green function whenever $x$ and $x'$ are
connected by a unique geodesic \cite{Hadamard, Poisson:2011nh}.

\begin{figure}[ht]
	\includegraphics[width=0.9\columnwidth]{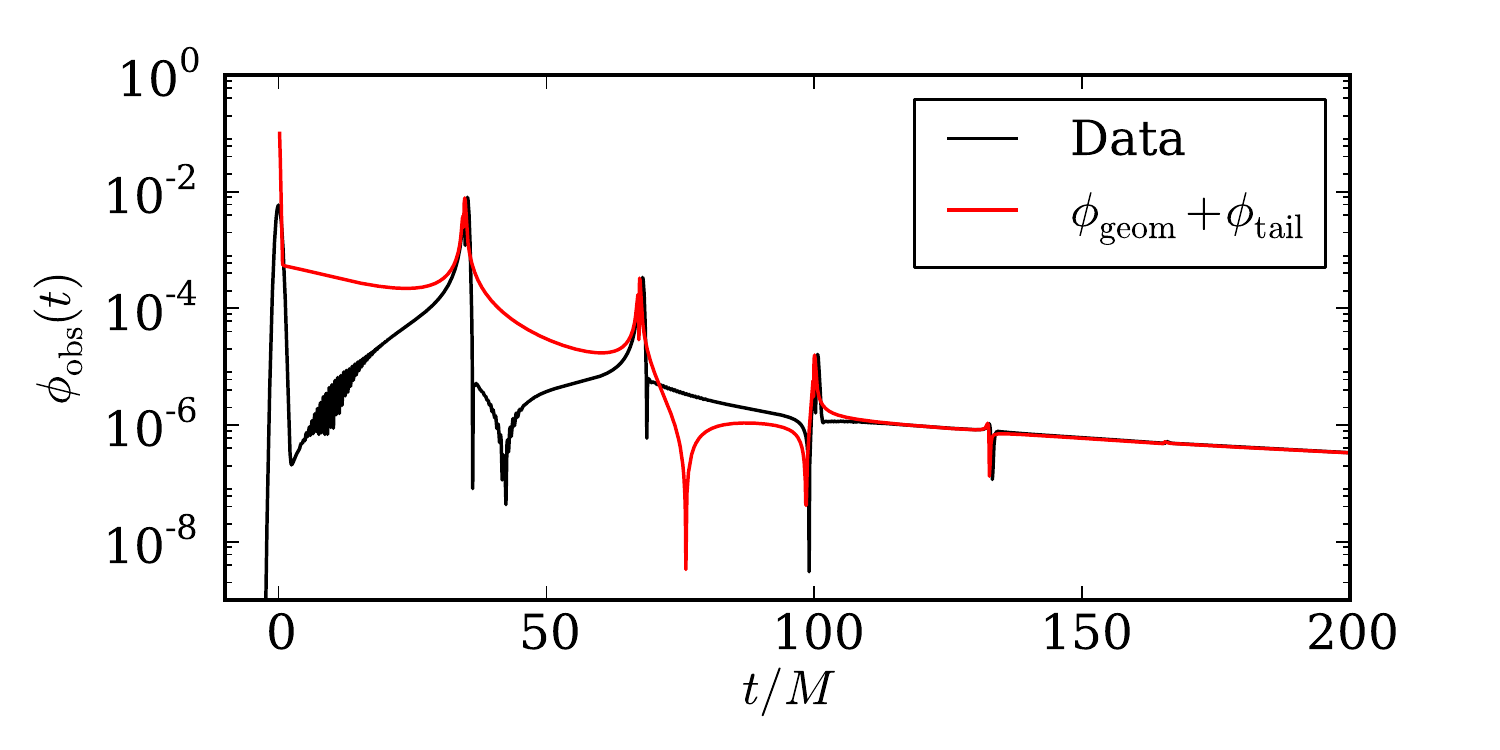}
	\caption{Reconstructing the numerical solution (black) as seen
          by an observer at null infinity on the positive
          $x$ axis along the caustic line. The analytical
          approximation (red) disagrees significantly with the data
          between the caustic echoes at early times.}
	\label{fig:divorciados}
\end{figure}

Our heuristic formula should fail along the caustic line (the $x$ axis
in our simulations) because the geometrical optics approximation
breaks down at caustics. We find that (\ref{eq:phigeom1}) and
(\ref{eq:phitail1}) on the caustic line works well for the
high-frequency part of the signal (i.e., the caustic echoes
themselves) and for describing the late-time tail (Fig.~\ref{fig:divorciados}), but cannot capture
the effects of early-time backscattering (compare to
Fig.~\ref{fig:enchilada} for the observer at $\gamma = \pi/2$). Also,
the formula has zero crossings which do not occur at the correct
times. Near a caustic the wavefronts are focused and magnified thus
distorting the shape of the wavepacket (see
Figs.~\ref{fig:3D_1}-\ref{fig:3D_4}). Hence, the echoes have a comparatively larger amplitude than the direct signal (Fig.~\ref{fig:divorciados}). These features are not accounted
for in the geometrical optics limit and require a more complete
description of wave optics in curved spacetime.

Further analysis is needed to improve on our heuristic formula. 
For example, one can include corrections to the geometrical optics
approximation using the expansions presented in \cite{Dolan:2009nk}
or combine other quasilocal expansions of the Green function 
\cite{Ottewill:2009uj} and its Pad\'{e} resummation \cite{Casals:2009xa}. 
These potential directions for research are left for future work. 

\section{Conclusions}
\label{sec:conclusions}

We numerically solved the scalar wave equation on a Schwarzschild
background \eqref{eq:sc} with a narrow Gaussian source
\eqref{eq:source} thus yielding an approximation to the Schwarzschild Green function
\eqref{eq:green}. For the numerical computations we used the Spectral
Einstein Code {\texttt SpEC} \cite{SpECWebsite}.

The main result of our study is the numerical approximation of the
full retarded Schwarzschild Green function in the time domain. 
This computation allowed us to demonstrate a cyclic singularity structure
due to the trapping of energy near the photon sphere.

The numerical evolution, visualized in
Figs.~\ref{fig:3D_1}-\ref{fig:3D_4} and in \cite{video}, proceeds as
follows. The narrow Gaussian source triggers a high-frequency
wavefront that propagates along null geodesics in accordance with
geometrical optics. Part of the energy of the initial wavefront gets
trapped at the black hole horizon and leaks out to infinity decaying
exponentially in time. At each half revolution around the photon
sphere the trapped wavefront forms a caustic resulting in
echoes that propagate out to infinity. The wavefront undergoes
a Hilbert transform at the caustics causing a $- \pi/2$ phase shift in
its profile, which results in a fourfold cycle of caustic echoes seen by generic observers. 
This cycle is known as the fourfold singularity structure of the retarded
Green function \cite{Ori, Casals:2009xa,Casals:2010zc,
  Dolan:2011fh,Casals:2012px,Harte:2012uw}.

The fourfold structure is a recent discovery in the context of
curved spacetimes but, in hindsight, it is a simple consequence of
well-known facts. The $\pi/2$ phase shift of wavefronts at caustics
has already been discovered in the late 19th century and is widely known as the Gouy phase shift in optics \cite{Gouy, Hartmann:2009}. 
Combined with the notion of trapping (e.g., 
due to the presence of a photon sphere), this knowledge
immediately implies a fourfold cycle in a spherically symmetric, asymptotically flat spacetime \footnote{Trapping may also
occur globally in cosmological spacetimes (e.g., the Einstein static universe), which may result in a
different phase shift and therefore a different $n$-fold
structure.}. The viewpoint of a phase shift of the Green function 
at caustics seems more fundamental than the fourfold structure. 
For example, a phase shift may be observed in regions of moderate
curvature due to gravitational lensing where no trapping occurs, 
whereas trapping is required for the observation of $n$-fold cycles.
In addition, a nongeneric set of observers, those on the caustic line, 
see a twofold cycle. Hence, the number of echoes in a cycle is 
observer-dependent.

With the full numerical approximation to the Green function at hand, 
we also obtained an interesting result regarding gravitational lensing.  
If backscatter off background curvature is included, only a finite number
of images are visible, even with ideal instruments, for any source of finite 
wavelength because caustic echoes decay exponentially in amplitude
while backscatter (typically neglected in lensing studies)
decays only polynomially.

Further results can be summarized as follows. 
The arrival and decay of
successive caustic echoes are consistent with the orbital period
and Lyapunov exponent of null geodesics trapped at the photon
sphere. The arrival times of successive half-period echoes depend on
the source-observer configuration, with degeneracy resulting in a
twofold cycle when the source and the observer are aligned with the
black hole. This is the caustic line along which the amplified signal
propagates. The energy magnification at the caustic
follows an inverse power law with the scale of the wavefront in
accordance with predictions going beyond the geometrical optics limit
\cite{KayKeller}.  The backscatter follows
well-known polynomial decay rates at null infinity and at finite
distances, which limits the number of echoes measurable by observers.

We formulated an analytic approximation to the Schwarzschild Green
function in the geometrical optics limit including wavefront
propagation through caustics. Combined with our quantitative
understanding of the dynamics, this formula allowed us to capture the
essential features of the observed signal in a heuristic
expression. Given the simplicity of our approximations, it is
remarkable that the expression agrees so well with the numerical
calculation.

The only sources of error in the numerical computation of the
Schwarzschild Green function are the finite width of the Gaussian
source \eqref{eq:source} and the truncation error due to
discretization of the scalar wave equation. The boundary error
typically present in such simulations is eliminated by hyperboloidal
scri-fixing in a layer \cite{Zenginoglu:2007jw,Zenginoglu:2010cq}.
We emphasize that hyperboloidal compactification is not necessary for
the numerical study of the Green function near the black hole. 
Many of our results are reproducible using
standard foliations. However, hyperboloidal compactification helps
focusing the resolution to the strong field domain without
contamination from artificial boundaries and without wasting computational
resources on the asymptotic domain. It allows us to compute the long-time
evolution accurately at low cost, and provides us direct
numerical access to measurements of an idealized observer at future
null infinity.

The error scales in our computation are related. To evolve a narrow
Gaussian source, we need a large number of collocation points for the
spectral expansion of the variables. Failure to do so results in Gibbs
phenomena and contaminates the evolution. (Some high frequency noise
in the late-time evolution can be seen in the lower curve of
Fig.~\ref{fig:tail}). Numerical convergence tests and the agreement
between theory and experiment indicate that our errors are small but
we did not present a detailed error analysis in this paper. Such an
analysis will be required when comparing the
numerical simulations to accurate approximations of the Green
function or when computing the self-force acting on a particle using
the numerically computed Green function. For these calculations,
horizon-source ratios larger than 10:1 might be necessary.  The
numerical method must be further improved for very high ratios,
possibly using a second order in space formulation
\cite{Taylor:2010ki}, implicit-explicit time stepping
\cite{Lau:2008fb}, and adaptive mesh refinement. There are also
adapted methods to solve for high-frequency wave propagation, such as
the frozen Gaussian beam method \cite{LuYang} or the butterfly
algorithm \cite{Butterfly}, that may improve the efficiency of the 
numerical simulation considerably.

An immediate extension of our study would be the numerical
computation of retarded Green functions in Kerr spacetime. In rotating
black hole spacetimes, frame dragging plays a
role in the wavefront propagation. There is no photon
sphere in Kerr spacetime; instead, there are spherical photon orbits
bounded by the location of pro- and retrograde circular photon orbits
\cite{Teo:2003} which may participate in the Cauchy evolution in a
similar way as does the photon sphere in Schwarzschild spacetime. 
The large $\ell$ limit of the quasinormal mode spectrum and its relation to 
spherical photon orbits in Kerr spacetime have recently been analyzed in
\cite{Yang:2012}. Such analytic knowledge can be combined with numerical experiments to reveal the structure of the Green function in Kerr spacetimes 
and to provide good approximations to it. 
The visualization of a numerical Kerr Green function simulation using {\tt SpEC} can be
found in \cite{videoKerr}.

An interesting application of numerical approximations of Green functions
would be the computation
of the self-force on a small compact object moving in a supermassive black hole spacetime. Note that the Green function provides a very general
way of solving for wave propagation in a background spacetime. If good
approximations to it can be found, possibly augmented by data from
numerical computations, the method can be applied to essentially any
problem on the given background via suitable convolutions. To
this end, it may be useful to have sparse representations of the numerical 
Green function.

The improvement of Green function approximations may benefit from
developments in various fields. Formation of caustics in black hole
spacetimes is an example of the emergence of discrete structures from
continuous ones, a hallmark of catastrophe theory \cite{Arnold}. A
development of geometric diffraction theory and wave optics for black
holes may allow us to compute the Green function through caustics
without matching arguments while also providing more accurate analytical approximations. Exciting developments in these directions
can be expected in the near future.

\acknowledgments 

We thank Marc Casals, Sam Dolan, Haixing Miao, Mark Scheel, B\'ela
Szil\'agyi, Nicholas Taylor, Dave Tsang, and Huan Yang
for discussions. AZ is supported by the NSF Grant No.~PHY-1068881, 
and by
a Sherman Fairchild Foundation Grant to Caltech. CRG is supported by
an appointment to the NASA Postdoctoral Program (administered by Oak Ridge Associated Universities through a contract with NASA) at the Jet Propulsion Laboratory, California Institute of Technology, under a contract with NASA where part of this research was carried out. Copyright 2012. All rights reserved.

\appendix

\section{Hyperboloidal layer with excision}
\label{app:layer}

In our numerical simulations we use hyperboloidal scri-fixing
\cite{Zenginoglu:2007jw} in a layer \cite{Zenginoglu:2010cq} attached
to a finite Schwarzschild domain in ingoing Eddington--Finkelstein
(iEF) coordinates. In this Appendix we describe the construction of
such coordinates.

The Schwarzschild metric in standard Schwarzschild coordinates 
$\{t_S,r,\vartheta,\varphi\}$ reads
\[ g =-\left(1-\frac{2M}{r}\right)\,dt_S^2 + \left(1-\frac{2M}{r}\right)^{-1} dr^2 + r^2 d\sigma^2,  \]
where $d\sigma^2 = d\vartheta^2 + \sin\vartheta^2 d\varphi^2$ is the standard metric on the unit sphere. 
Ingoing Eddington--Finkelstein coordinates are constructed from
Schwarzschild coordinates by the time transformation $t =t_S + 2M \ln
(r-2M)$, which leads to the metric
\begin{eqnarray}\label{eqn:iEFmetric} 
g &=& -\left(1-\frac{2M}{r}\right)\,dt^2 + \frac{4M}{r}\, dt \, dr
\nonumber \\ && + \left(1+\frac{2M}{r}\right)\, dr^2 + r^2 \,
d\sigma^2. \end{eqnarray} 
The metric is regular at the horizon, $r=2M$, because the slices of
the iEF foliation penetrate the future horizon without intersecting each
other, as opposed to the slices of the standard Schwarzschild foliation
which intersect at the bifurcation sphere. In a similar construction, iEF
slices that intersect at spatial infinity can be transformed to
hyperboloidal slices that foliate null infinity
\cite{Zenginoglu:2011jz}.  The hyperboloidal layer method that leads
to such a slicing consists in the introduction of a compactifying
coordinate in combination with a suitable time transformation. The
compactifying coordinate $\rho$ can be defined via 
\be \label{eq:space} r = \frac{\rho}{\Omega}, \qquad \rm{where} \qquad  \Omega = 1- \left(\frac{\rho- \it{I}}{\it{S} - \it{I}}\right)^4\Theta(\rho - \it{I})\,.\ee 
Here, $\Theta$ is the Heaviside step function. The compactifying layer
is attached at the interface $\rho=I$ and has a thickness of $S-I$ in
local coordinates. The zero set of $\Omega$, namely $\rho=S$,
corresponds to infinity. To avoid loss of resolution near infinity, we
combine this spatial compactification with a time
transformation. The time transformation is constructed requiring invariance of  outgoing 
characteristic speeds 
\cite{Zenginoglu:2010cq, Jasiulek:2011ce}, or equivalently, of the outgoing null surfaces in local coordinates \cite{Bernuzzi:2011aj, Zenginoglu:2011zz}. Outgoing null
surfaces satisfy
$ u = t - \left(r + 4M \log(r-2M)\right). $
The invariance condition translates to
\[  t - \left(r + 4M \log(r-2M)\right) =  \tau - \left(\rho + 4M \log(\rho-2M)\right).\]
This relation defines the hyperboloidal time coordinate $\tau$ used in
the compactifying layer. The metric resulting from these
transformations is singular up to a rescaling. The conformally
rescaled metric reads
\begin{widetext}
\begin{eqnarray*} \tilde{g} &=& \Omega^2 g =  -\Omega^2 \left(1-\frac{2M\Omega}{\rho}\right) d\tau^2 - 2\left(L-\Omega^2 \frac{(\rho+2M)(\rho-2M\Omega)}{\rho(\rho-2M)}\right) d\tau d\rho \\
&& + \frac{\rho+2M}{\rho(\rho-2M)^2} \left(2L\rho(\rho-2M)-\Omega^2
  (\rho+2M)(\rho-2M\Omega)\right) d\rho^2 + \rho^2 d\sigma^2 ,
\end{eqnarray*}
\end{widetext}
where $L\equiv\Omega - \rho \, d\Omega/d\rho$. The metric looks rather
complicated but notice that for $\Omega=1$ (or equivalently, for
$\rho<I$) it reduces to the iEF metric \eqref{eqn:iEFmetric} and at
null infinity, $\Omega=0$, it takes a purely outgoing form. As a
consequence, characteristic speeds incoming with respect to the bulk
vanish at the boundaries.

\begin{figure}[ht]
\center \includegraphics[width=\columnwidth]{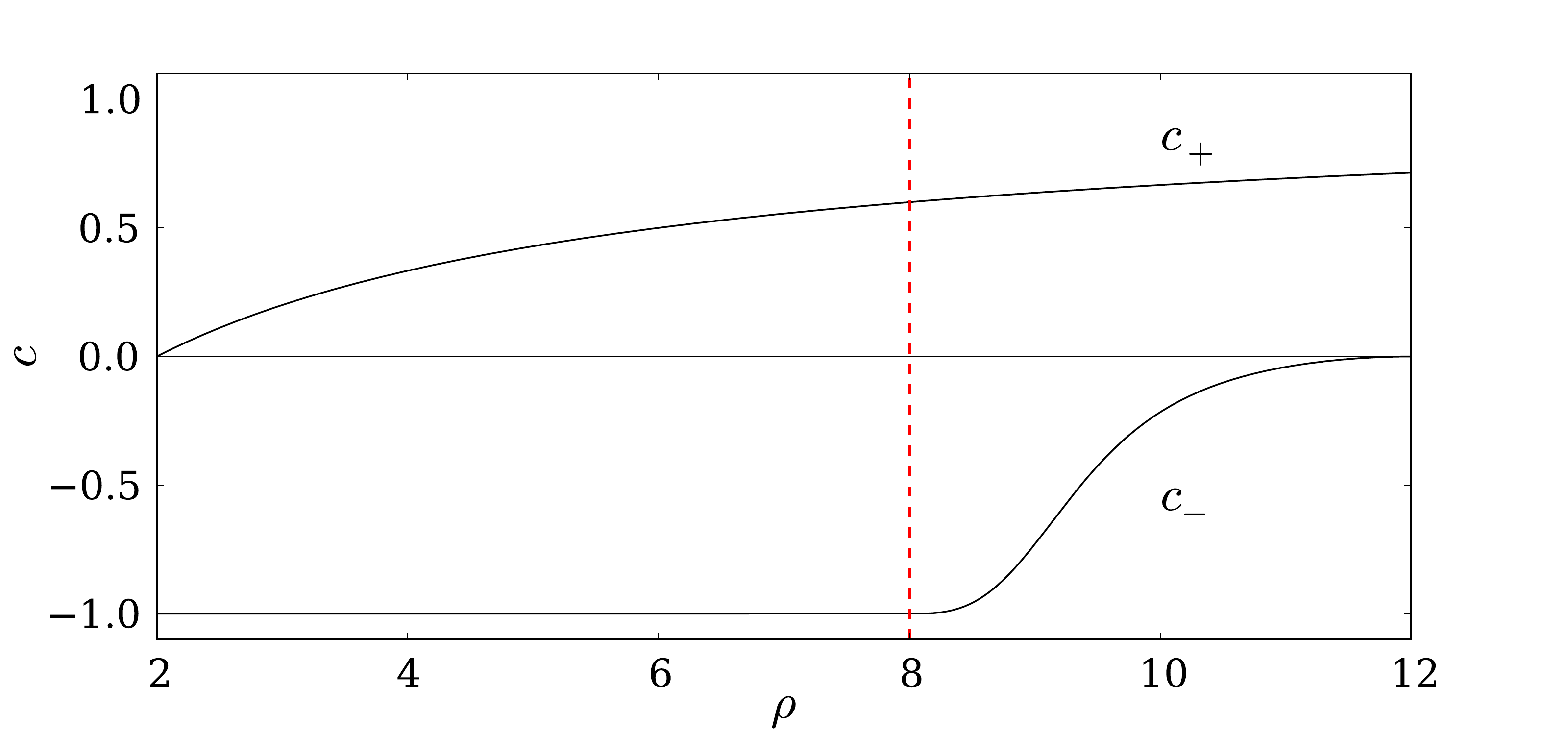}
\caption{Radial characteristic speeds in ingoing
  Eddington--Finkelstein coordinates with a hyperboloidal layer
  attached. The dashed vertical red line depicts the interface to the
  layer at $I=8$. Null infinity is at the coordinate location
  $S=12$. These parameters are also used in the numerical simulations.
\label{fig:iEFsps}}
\end{figure}

\begin{figure}[ht]
\includegraphics[width=0.93\columnwidth]{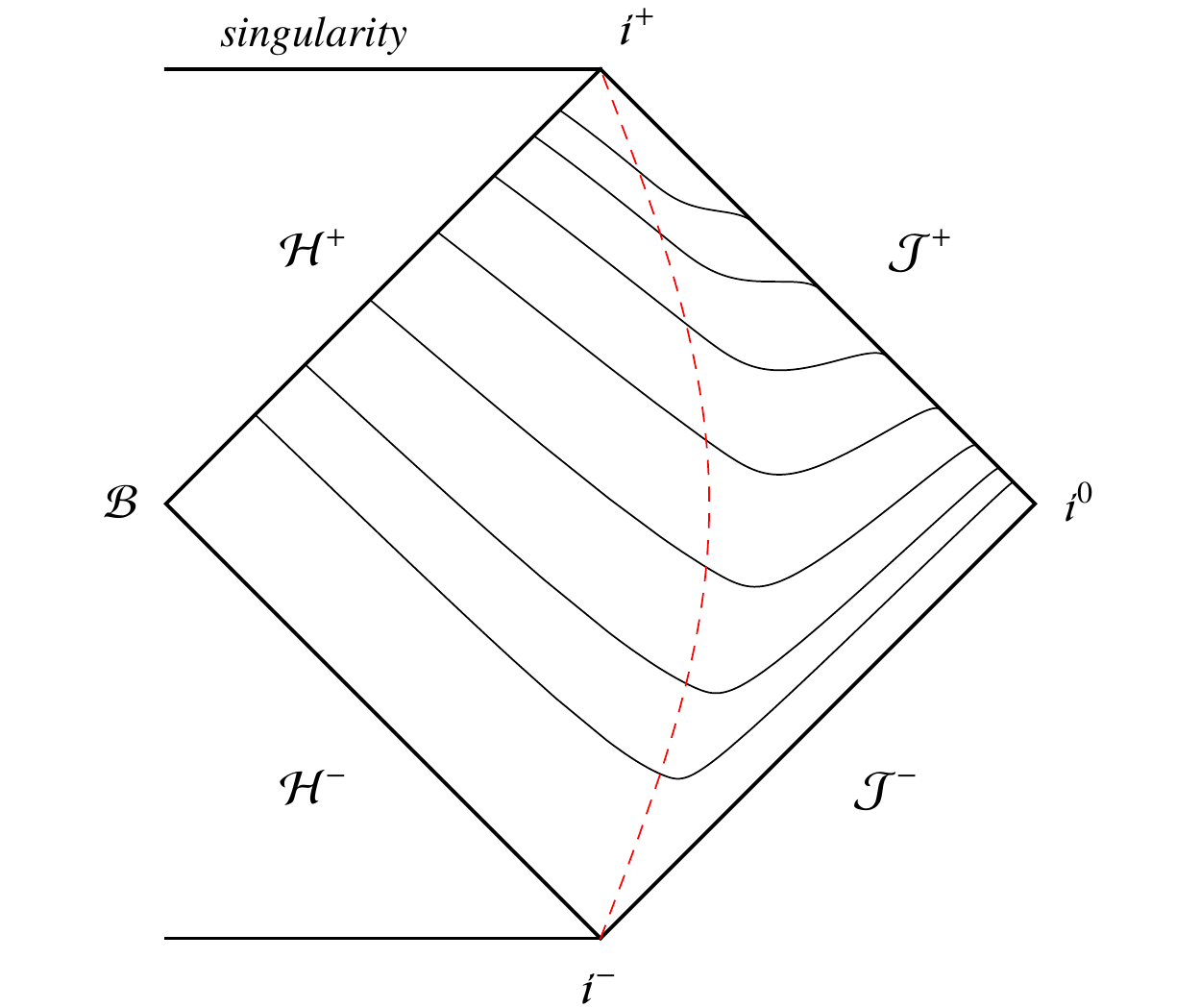}
\caption{Conformal diagram of an ingoing Eddington--Finkelstein
  foliation with a hyperboloidal layer. The dashed red line
  corresponds to the interface. For visualization we have chosen $I=3$
  and $S=4$. For the numerical computations we set $I=8$ and $S=12$ as
  in Fig.~\ref{fig:iEFsps}.
\label{fig:iEFconf}}
\end{figure}

The radial characteristic speeds read
\[ c_+ =  \frac{\rho-2M}{\rho+2M}, \] 
\[ c_- = - \Omega^2 \frac{(\rho-2M)(\rho-2M\Omega)}{2L\rho(\rho-2M)+\Omega^2(\rho+2M)(\rho-2M\Omega)} .\]
The outgoing speed $c_+$ has the same form as in iEF
coordinates. Consequently, it vanishes at the event horizon
$r=\rho=2M$. Similarly, the ingoing speed $c_-$ vanishes at null
infinity, $\rho=S$ where $\Omega = 0$. Figure \ref{fig:iEFsps} shows
the radial characteristic speeds in these coordinates. The conformal
diagram of the iEF foliation with a hyperboloidal layer is depicted in
Fig.~\ref{fig:iEFconf}.

In the hyperboloidal layer we solve the conformally transformed 
scalar wave equation \eqref{eq:sc} that reads \cite{Wald84}
\be\label{eq:confsc} \Box\Phi-\frac{1}{6} R \, \Phi = \Omega^{-3}
S(x),\ee where $\Phi = \phi/\Omega$ and $R$ is the Ricci scalar of the
conformal background. The divisions by the conformal factor do not
pose any difficulty because the scalar field falls off suitably and
the source has compact support.

\section{Green function in geometrical optics}
\label{app:optics}

The numerical solutions presented in Sec.~\ref{sec:results} can be
regarded as approximations to the Green function for the scalar wave
equation (\ref{eq:sc}).  We derive here, for the first time to our
knowledge, the scalar Green function in Schwarzschild spacetime in the
geometrical optics limit, which is used in the main text to compare
numerical data with analytical approximations and to explain the
fourfold cycle and the profiles of the caustic echoes. See \cite{Hartmann:2009}
for a review of the geometrical optics limit in the field of optics.

In Schwarzschild spacetime there exists a time-like Killing vector,
$\partial_t$, that allows for the Green function to be decomposed into
frequency modes via
\begin{align}
	G(x, x') = \int_{-\infty}^\infty \frac{ d\omega }{ 2\pi } \, e^{ - i \omega (t-t') } G (\omega; {\bf x}, \bf{x}' ) .
\label{eq:GreenFD}
\end{align}
In the frequency domain, using coordinates in which $g^{ti}=0$, the Green
function equation (\ref{eq:green}) becomes a Helmholtz equation
\begin{align}
	\Delta G(\omega; {\bf x}, {\bf x}') + \omega^2 u^2( {\bf x} ) G(\omega; {\bf x}, {\bf x}') = 0,
\label{eq:helmholtz1}
\end{align}
where $\Delta \equiv g^{ij} \nabla_i \nabla_j$ is the curved space Laplacian (with $i,j=1,2,3$), and we assume
that ${\bf x} \ne {\bf x}'$ so that the Dirac delta distribution in
(\ref{eq:green}) vanishes. The function $u({\bf x})$ is defined by
\begin{align}
	u^2( {\bf x}) = - g^{tt} ({\bf x}).
\end{align}
In standard Schwarzschild coordinates $u^2(\bx) = 1- \tfrac{2M}{r}$.

In the geometrical optics limit, the frequency of the mode approaches
infinity (the wavelength of the mode is much smaller than the
background curvature scale).  We make the following ansatz for the
asymptotic expansion of the Green function,
\begin{align}
	G(\omega ; \bx, \bx' ) \sim \sum_{n=0}^\infty \frac{ A_n (\bx, \bx') }{ (i \omega)^n } \, e^{i \omega T(\bx, \bx') }  .
\label{eq:ansatz1}
\end{align}
Here, $T( \bx, \bx')$ is called the {\it eikonal} and its level
surfaces are surfaces of constant phase, called {\it wavefronts}. The
covector $\nabla_i T$ is everywhere normal to this surface. Note that
this ansatz is not covariant as time is a preferred direction. This
may seem undesirable but our static observers measuring the field, as
in Fig.~\ref{fig:logplot}, are Killing observers. Additionally, the
geometrical optics approximation emphasizes time over space via the
assumption of only high-frequencies contributing to the field. Hence, the
lack of manifest covariance is suitable for our purposes.

\begin{figure}
	\includegraphics[width=0.75\columnwidth]{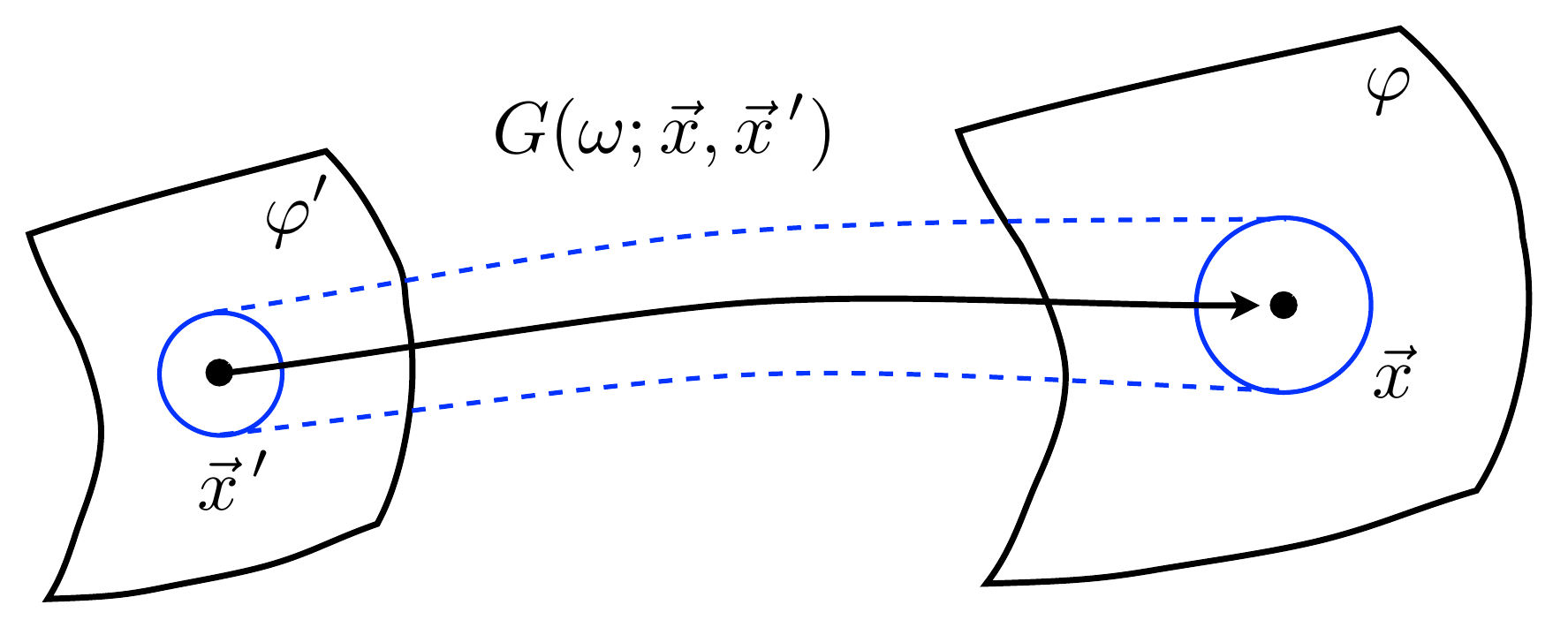}
	\caption{The retarded Green function as a mapping
          from a wavefront with phase $\varphi'$ to a
          wavefront with phase $\varphi > \varphi'$.}
	\label{fig:propagator}
\end{figure}

Different level surfaces correspond to wavefronts with different
phases of the Green function. We may thus parameterize these
wavefronts by the phase, which is an affine parameter denoted by
$\varphi$. As the field increases phase, and thus evolves forward in
time, one moves from one wavefront to the next.  This motion can be
regarded as a mapping that allows one to identify paths, or {\it
  rays}, that connect specific points on different wavefronts.  The
Green function in (\ref{eq:ansatz1}) thus causally propagates points
from one wavefront to another (Fig.~\ref{fig:propagator}). This
viewpoint will be useful below.

Substituting the leading order term of the sum in (\ref{eq:ansatz1})
into (\ref{eq:helmholtz1}) and equating like powers of $\omega$ gives
the following set of equations for the eikonal $T(\bx, \bx')$ and the
leading order amplitude $A_0 (\bx, \bx')$,
\begin{align}
	\nabla_i T \nabla^i T &= u^2 (\bx) 
\label{eq:eikonal1} \\
	 2  \nabla_i A_0 \nabla^i T & = - A_0 \Delta T  .
\label{eq:transport1}
\end{align}
The first equation is called the {\it eikonal equation} and the second
is called the {\it transport equation}.  We will solve these equations
in turn.

The solution to the eikonal equation (\ref{eq:eikonal1}) can be found
by first defining a momentum vector $p_i \equiv \nabla_i T(\bx)$,
which is normal to the wavefront so that the eikonal equation reads
$\bp^2 = u^2(\bx)$. This defines a Hamiltonian for the rays connecting
one wavefront to another
\begin{align}
	H = \frac{1}{2} \big( p_i p_j g^{ij} (\bx) - u^2 (\bx) \big)  .
\label{eq:hamiltonian1}
\end{align}
The Hamiltonian vanishes when the eikonal equation is
satisfied. Hamilton's equations give
\begin{align}
	\frac{ d x^i }{ d\varphi} & = p^i = \nabla^i T(\bx) 
\label{eq:hamiltonseq1} \\
	\frac{ D p_i }{ d\varphi } & = \frac{1}{2} \nabla_i u^2 (\bx)
\label{eq:hamiltonseq2}
\end{align}
where $D/d\varphi = (dx^i/d\varphi) \nabla_i$ is the covariant
parameter derivative along a trajectory with coordinates $x^i
(\varphi)$ (i.e., a {\it ray}). Solving these ray equations thus
amounts to solving the eikonal equation.

The square of the ray's velocity is
\begin{align}
	\frac{ dx^i }{ d\varphi } \frac{ dx_i }{ d\varphi } = \nabla_i T \nabla^i T = u^2 (\bx) = -g^{tt}(\bx)  .
\end{align}
Collecting all terms on one side and multiplying by $d\varphi^2$ yields
\begin{align}
	g^{tt}(\bx) d\varphi^2 + g^{ij} (\bx) dx_i dx_j = 0  .
\end{align}
Identifying $dx^0 = dt = u^2(\bx) d\varphi$, we see that the rays in the geometrical optics limit
are null because $dx^\mu dx_\mu = 0$.  The solution to the eikonal
equation thus amounts to solving for the trajectories of the null
rays, which is equivalent to solving the geodesic equation for two
light-like separated points, one on each wavefront.

Next, we solve the transport equation (\ref{eq:transport1}). Dividing
both sides by $A_0$, noting that $\Delta T = \nabla_i
\dot{x}^i(\varphi)$, and $\nabla^i T \nabla_i = D / d\varphi$ (from
(\ref{eq:hamiltonseq1})) yields
\begin{align}
	\frac{ d \ln A_0^{-2} }{ d\varphi }  = \nabla_i \dot{x}^i  .
\label{eq:transport2}
\end{align}
If an infinitesimal area on the wavefront centered at $\bx' \equiv
\bx(\varphi')$ is evolved along the null geodesics to phase $\varphi >
\varphi'$ then the subsequent evolution can be viewed as a change of
variable since $\bx \equiv \bx(\varphi)$ depends on the ``initial
data'' at $\bx'$ and implies the coordinate transformation $\bx = \bx
(\varphi; \bx')$. The Jacobian relating these areas is
\begin{align}
	J ( \bx, \bx' ) = \frac{ \partial ( x ^0 \cdots x^3) }{ \partial ( x'{} ^0 \cdots x'{}^3) } = {\rm det} M = e^{ {\rm Tr} \ln M},
\label{eq:jacobian1}
\end{align}
where the matrix $M$ has components $M^i{}_j = \partial x^i / \partial
x'{}^j$. Taking the parameter derivative of the Jacobian gives
\begin{align}
	\frac{ d J }{ d\varphi} = {} &  J \frac{ \partial x'{}^i }{ \partial x ^j } \frac{ \partial \dot{x} ^j }{ \partial x'{} ^i } = J \frac{ \partial x'{}^i }{ \partial x^j } \frac{ \partial \dot{x}^j }{ \partial x^k } \frac{ \partial x^k }{ \partial x'{} ^i}  = J \frac{ \partial \dot{x}^i }{ \partial x^i } .
\end{align}
Using $\partial_i \dot{x}^i = \nabla_i \dot{x}^i - \Gamma^i_{i\alpha} \dot{x}^\alpha$ and
\begin{align}
	\Gamma^i _{i\alpha} \dot{x}^\alpha = {} & \Gamma^i _{i t} + \Gamma^i _{ij} \dot{x}^j = \dot{x}^j \partial_j \ln \sqrt{ h } \nonumber \\
	= {} & \frac{ d }{ d\varphi } \ln \sqrt{ h } 
\end{align}
where $h$ is the
determinant of the spatial metric $g_{ij}$, we obtain from
(\ref{eq:transport2})
\begin{align}
	\frac{ d \ln A_0^{-2} }{ d\varphi} = \nabla_i \dot{x}^i = \frac{ d }{ d\varphi} \big( J \sqrt{ h }  \big)  .
\end{align}
This has the solution
\begin{align}
	A_0 (\bx, \bx') = A_0 (\bx', \bx'') \left[ \frac{J(\bx', \bx'') }{ J (\bx, \bx')} \frac{ \sqrt{ h (\bx') }  }{  \sqrt{ h (\bx) }  }  \right]^{1/2},  
\label{eq:transport3}
\end{align}
provided that $\bx = \bx(\varphi)$ has not passed through any caustics
and where $A_0 (\bx', \bx'')$ is initial amplitude ($\bx''$ is either
a point on an earlier wavefront or can be equal to $\bx'$ in which
case $J(\bx', \bx') = 1$).  Hence, the amplitude of the wave at phase
$\varphi'$ for a given infinitesimal area around the point $\bx' =
\bx(\varphi')$ on the wavefront will evolve along the null ray to $\bx
= \bx(\varphi)$ with phase $\varphi > \varphi'$ according to
(\ref{eq:transport3}). Notice that the amplitude at $\bx$ depends on
the expansion or contraction of the infinitesimal area element on the
wavefront during its evolution. This is equivalently described by the
expansion or contraction of a congruence of null geodesics connecting
$\bx'$ and $\bx$. When the ray passes through a caustic the Jacobian
$J(\bx, \bx')$ changes sign relative to $J(\bx', \bx'')$. We will say
more about this in Appendix \ref{app:hilbert}.

Assuming that the ray has not passed through any caustics, we may
reconstruct the time-domain Green function in (\ref{eq:GreenFD}) using
(\ref{eq:ansatz1}) and (\ref{eq:transport3}) to find at leading order
in $1/\omega$,
\begin{align}
	G( x, x') \sim {} & A_0 (\bx, \bx') \left[  \frac{J(\bx', \bx'') }{ J (\bx, \bx')} \frac{ \sqrt{ h (\bx') }  }{  \sqrt{ h (\bx) } } \right]^{1/2} \nonumber \\
		& \times {\rm Re} \int_{-\infty}^\infty \frac{ d\omega }{ 2\pi } e^{ - i \omega (t - t' - T(\bx, \bx' ) ) },
\label{eq:N0green0}
\end{align}
where we take the real part of the integral because the field in the
time domain is real valued. The evaluation of the integral gives
\begin{align}
	G( x, x' ) \propto \delta \big( t - t' - T( \bx, \bx') \big)  .
\label{eq:N0green}
\end{align}
Hence, the geometrical optics approximation yields a contribution to
the Green function whenever the time difference between $x'$ and $x$
satisfies $t - t' = T( \bx, \bx')$, which is just the time it takes
for a null geodesic to connect these two points.

Finally, we note that the remaining amplitudes $A_n (\bx, \bx')$ in
(\ref{eq:ansatz1}) can be determined through a recursion relation (for
$n \ge 0$),
\begin{align}
	2 \nabla^i T \nabla_i A_{n+1} + A_{n+1} \Delta T = - \Delta A_n  ,
\end{align}
which follows from substituting (\ref{eq:ansatz1}) into
(\ref{eq:helmholtz1}) and setting the coefficients of each power of
$\omega$ to zero. With $A_0$ known from (\ref{eq:transport3}) and the
rays found by solving the null geodesic equation, we have the Schwarzschild Green
function in the geometrical optics limit. The rays are null at all
orders of $1/\omega$. Hence, the geometrical optics limit is
inapplicable to the propagation {\it into} the null cone of $\bx'$ and
thus ignores backscattering off the background curvature.

\section{Hilbert transform through a caustic}
\label{app:hilbert}

\begin{figure}[ht]
	\includegraphics[width=\columnwidth]{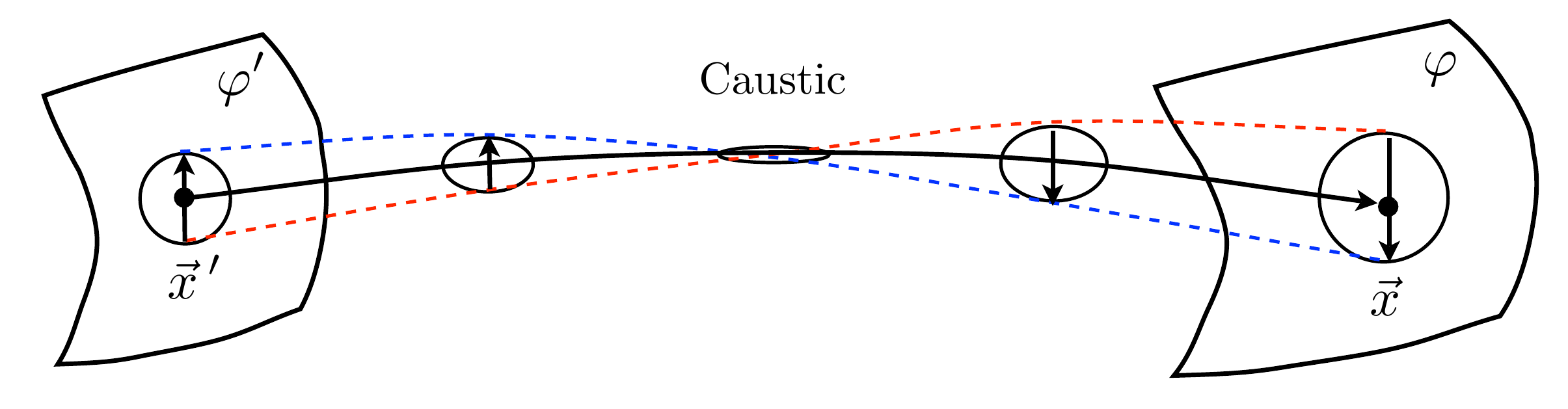}
	\caption{A cartoon of the change in orientation of an
          infinitesimal area element on the wavefront's spatial
          section as it passes through a caustic in Schwarzschild spacetime. Only one
          dimension of the area element is compressed.}
	\label{fig:thrucaustic}
\end{figure}

In Appendix \ref{app:optics} we solved the transport equation
(\ref{eq:transport1}) for the field's amplitude in the geometrical
optics limit. The solution (\ref{eq:transport3}) assumes that the
quantity under the radical is positive. However, if $\bx =
\bx(\varphi)$ evolves through a caustic then the Jacobian
(\ref{eq:jacobian1}) changes sign. We show in
Fig.~\ref{fig:thrucaustic} a cartoon of a wavefront area element as it
passes through a generic caustic.

The sign change of the Jacobian through a caustic implies a purely
imaginary amplitude in (\ref{eq:transport3}). Hence, the amplitude
changes phase by $\pm \pi/2$ through a caustic \cite{Gouy, Hartmann:2009}. The sign of the phase
shift is determined by matching to the full solution from numerical
simulations. It cannot be computed within the geometrical optics
approximation because the latter breaks down at caustics (where
$J(\bx, \bx') = 0$ and $A(\bx, \bx')$ becomes singular).

After some trial and error, we find that the phase of
(\ref{eq:transport3}) when passing through one caustic is given by
\begin{align}
	\exp \left( - \frac{i \pi }{ 2 } {\rm sgn} ( \omega) \right)  .
\label{eq:phase1}
\end{align}
For positive frequencies the phase is $ -i$ while for negative
frequencies it is $ +i$. The frequency-dependence in
(\ref{eq:phase1}) is needed for the reconstruction of the time-domain
Green function in the geometrical optics limit. Computing
(\ref{eq:GreenFD}), upon substituting the leading order term of the
$1/\omega$ expansion from (\ref{eq:ansatz1}), gives
\begin{align}
	G ( x, x') \sim {} & \big| A_0 (\bx', \bx'') \big| \left[ \bigg| \frac{ J(\bx', \bx'') }{ J(\bx, \bx') } \bigg| \sqrt{ \frac{ h (\bx') }{  h (\bx) } } \, \right]^{1/2}  \nonumber \\
		& \times \theta \big( t- t' - T_0 (\bx, \bx') \big) \nonumber \\
		&  \times {\rm Re} \int_{-\infty}^\infty \frac{ d\omega }{ 2\pi } \, e^{-i \omega ( t - t' - T_1(\bx, \bx') ) } e^{ - \frac{i \pi }{ 2 } {\rm sgn} (\omega) }, 
\label{eq:N1green}
\end{align}
where we have explicitly taken the real part as in
(\ref{eq:N0green0}), $T_N( \bx, \bx')$, with $N=0,1$, is the eikonal for the null
geodesic connecting $\bx$ and $\bx'$ and passing through $N$ caustics,
and the step function ensures causality (namely, that the $T_1$
eikonal is larger than $T_0$). The frequency integral gives
\begin{align}
	- \frac{1}{ \pi \big( t - t' - T_1(\bx, \bx' ) \big) }  .
\label{eq:oneoversigma1}
\end{align}
As discussed in Sec.~\ref{sec:hilbert}, when the Green function is
convolved with the source $S(x)$ from (\ref{eq:source}) one finds that
passing through one caustic transforms the field from a Gaussian
profile to that of a Dawson's integral (or ``Dawsonian''), which is the convolution of
(\ref{eq:oneoversigma1}) with the source $S(x)$
\cite{Duoandikoetxea}. Hence, the choice given in (\ref{eq:phase1})
for the sign of the root of $A_0^2 (\bx, \bx')$ in
(\ref{eq:transport3}) is justified through this matching calculation.

Comparing (\ref{eq:oneoversigma1}) to (\ref{eq:N0green}) we note that
the latter has support only along the null ray connecting $\bx$ and
$\bx'$ while the former has a much broader domain of support. Both
expressions are singular on null rays (where $t-t' = T_0(\bx, \bx')$ or $T_1 (\bx,
\bx')$), but (\ref{eq:oneoversigma1}) does not vanish in the forward
light cone thus adding to the tail. Passing through one caustic
causes the sharp delta distribution to become smeared out into the
future null cone of $\bx'$ beyond the caustic.

\begin{figure}[ht]
	\includegraphics[width=\columnwidth]{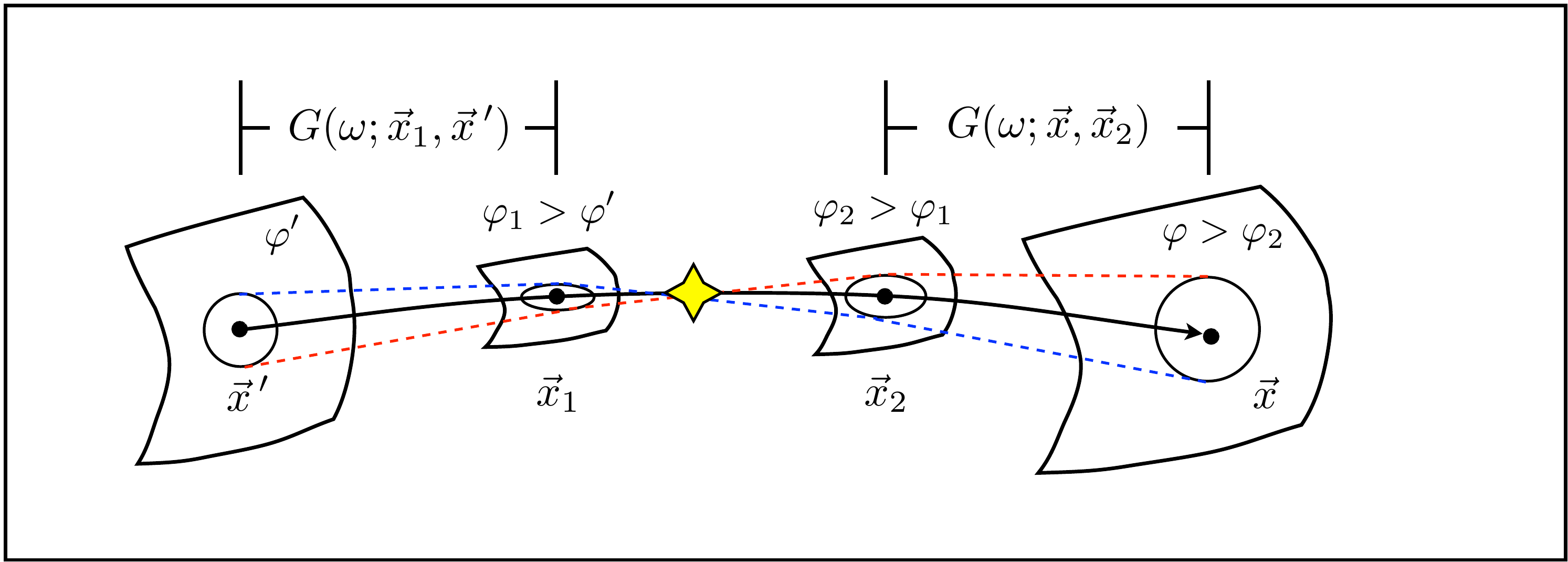}
        \\~\\ \includegraphics[width=\columnwidth]{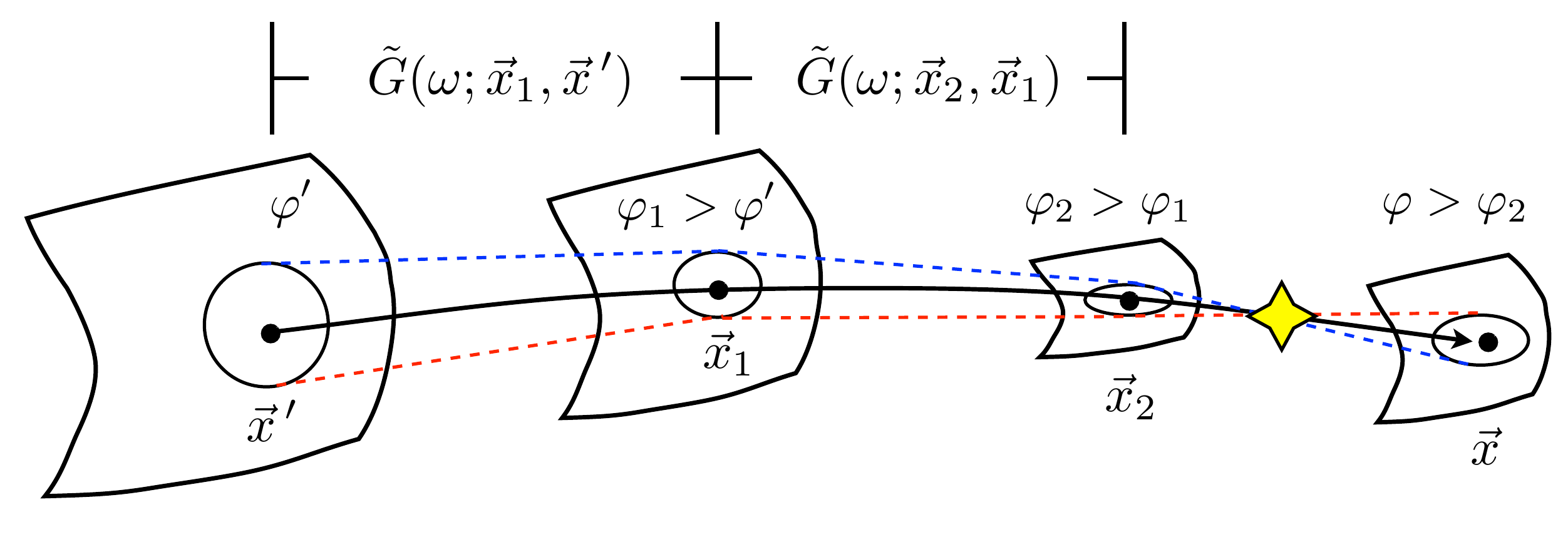}
	\caption{{\bf Top}: The retarded Green function in the
          geometrical optics limit for a null ray connecting $\bx'$ to
          $\bx$ through a single caustic. The
          geometrical optics approximation breaks down at the caustic 
          (yellow star) but is
          valid from the wavefronts with phase $\varphi'$ to
          $\varphi_1 > \varphi'$ and from the wavefront with phase
          $\varphi_2$ to $\varphi > \varphi_2$. {\bf Bottom}: In
          Maslov's theory the caustic is displaced by a phase space
          canonical transformation yielding a related Green function
          that allows for propagating the wavefront with phase
          $\varphi'$ directly to $\varphi_2$ via $\varphi_1$.}
	\label{fig:propagator2}
\end{figure}

The frequency integral in (\ref{eq:N1green}) is the {\it Hilbert
  transform} of $\delta (t - t' - T_1( \bx, \bx'))$. For a function
$f(t)$, the Hilbert transform in the frequency domain is defined as
\begin{align}
	H [ f(t) ] = \int_{-\infty}^\infty \frac{ d\omega}{2\pi} \, \hat{f} (\omega) e^{ -i \pi / 2 \, {\rm sgn} (\omega) },
\label{eq:hilberttransform1}
\end{align}
where $\hat{f}(\omega)$ is the Fourier transform of $f(t)$.  The
effect of the Hilbert transform is to shift the phase of
$\hat{f}(\omega)$ by $-\pi/2$. Two applications of the Hilbert
transform gives $H_2 [f(t) ] \equiv H [ H[ f(t) ] ] = - f(t)$, which
is a $- \pi$ phase shift of $f(t)$. Four applications of the Hilbert
transform gives back the function itself,
\begin{align}
	H_4 [ f(t) ] = f(t)  .
\end{align}
Thus, the Hilbert transform explains the fourfold cycle of the
retarded Green function through caustics.  This allows us to generalize
(\ref{eq:N1green}) to include {\it all} the null rays that connect
$\bx$ and $\bx'$ (of which there are infinitely many for a
Schwarzschild black hole \cite{Hasse:2001by}),
\begin{align}
	G (x, x') \sim {} & \sum_{N=0}^\infty \big| A_0 (\bx', \bx'') \big| \left[ \bigg| \frac{ J(\bx', \bx'') }{ J(\bx, \bx') } \bigg| \sqrt{ \frac{ h (\bx') }{  h (\bx) } } \, \right]^{1/2}  \nonumber \\
		 & \times  \theta ( t - t' - T_{N-1} (\bx, \bx') ) \, {\rm Re} \int_{-\infty}^\infty \frac{ d\omega }{ 2\pi } \nonumber \\
		&  {\hskip0.25in} \times e^{-i \omega ( t-t' - T_N(\bx, \bx') ) } e^{ - \frac{i \pi N}{ 2 } {\rm sgn} (\omega) }  ,
\label{eq:Ngreen}
\end{align}
where $T_{-1} (\bx, \bx') \equiv 0$.
This formula can be written more succinctly as
\begin{align}
	G( x, x' ) \sim {} &  \sum_{N=0}^\infty \big| A_0 (\bx, \bx') \big| \, \theta ( t - t' - T_{N-1} (\bx, \bx') ) \nonumber \\
		& {\hskip0.25in} \times {\rm Re} \big\{ H_N \big[ \delta ( t - t' - T_N (\bx, \bx') ) \big]  \big\},
\label{eq:Ngreencompact}
\end{align}
where $T_N (\bx, \bx')$ is the eikonal for the null geodesic that
passes through $N$ caustics and $H_N$ denotes the Hilbert transform
applied $N$ times to its argument.

As mentioned at the beginning of this Appendix, 
the geometrical optics approximation
breaks down at caustics where many points of a wavefront get mapped to
the same (caustic) point. The Jacobian $J(\bx, \bx')$ vanishes and the
amplitude of the field $A_0 (\bx, \bx')$ diverges as $J(\bx,
\bx')^{-1/2}$.  The geometrical optics approximation is valid only
between successive caustics. One may use the geometrical optics Green
function to evolve wavefronts before and after a caustic but not
through (Fig.~\ref{fig:propagator2}).

Remarkably, the breakdown of geometrical optics at a caustic can be
averted (or rather, displaced) using Maslov's theory
\cite{Maslov:book1965, Maslov:book1972, Kravtsov:1968}. A caustic
arises due to singularities in the projection of some higher
dimensional space to a lower one. In our case, the higher dimensional
space is the phase space of null rays (with coordinates $(\bx, {\bf
  p})$) wherein the null rays follow trajectories with coordinates
$(\bx(\varphi), {\bf p} (\varphi))$ that do not intersect because their
evolution is Hamiltonian [see (\ref{eq:hamiltonian1})].

Instead of forming the asymptotic series in $1/\omega$ for $G(\omega;
\bx, \bx')$, Maslov's idea is to form the series for its Fourier
transform with respect to at least one of its spatial coordinates.
Then the Fourier transform operates as a canonical transformation in
phase space. When the new momentum variables are projected out,
caustics form at different locations than previously.  This change of
caustic location allows us, in principle, to construct a well-defined
asymptotic series for the geometrical optics approximation of the
Green function by using Maslov's theory whenever a caustic is about to
be encountered (Fig.~\ref{fig:propagator2}).  In this way, one may
be able to derive directly the phase shift through a caustic in (\ref{eq:phase1})
without recourse to a matching argument. The application of Maslov's
theory is outside the scope of this paper.


\bibliography{refs}

\end{document}